\documentclass[a4paper,10pt,preprintnumbers,floatfix,superscriptaddress,aps,pra,twocolumn,showpacs,notitlepage,longbibliography,nofootinbib]{revtex4-2}

\usepackage[dvipsnames]{xcolor}
\usepackage[utf8]{inputenc}
\usepackage{amsmath}
\usepackage{amsfonts}
\usepackage{amssymb}
\usepackage{amsthm}
\usepackage{pgfplots}
\usepackage{graphicx}
\usepackage{lmodern}
\usepackage{sidecap} 
\usepackage{cancel}
\usepackage{physics}
\usepackage{mathtools}
\usepackage{dsfont}
\usepackage{comment}
\usepackage{subfig}
\usepackage{ragged2e}

\usepackage[colorlinks]{hyperref}
\hypersetup{
    colorlinks  = true,
    citecolor   = PineGreen,
    linkcolor   = MidnightBlue,
    urlcolor    = PineGreen
}

\graphicspath{{figures/}{./}}

\newtheorem{lem}{Lemma}

\newtheorem{corr}{Corollary}

\newtheorem{prop}{Proposition}

\def\tr{\operatorname{tr}}

\def\c2{\ensuremath{C^{(2)}}}
\definecolor{DeepTeal}{RGB}{0,102,120}

 \pgfplotsset{compat=1.18}

\begin{document}

\title{Optimising the relative entropy under semidefinite constraints}
\date{\today}
\author{Gereon Koßmann}
\affiliation{Institute for Quantum Information, RWTH Aachen University, Aachen, Germany}
\author{René Schwonnek}
\affiliation{Institut f\"{u}r Theoretische Physik, Leibniz Universit\"{a}t Hannover, Germany}
\begin{abstract}
Finding the minimal relative entropy of two quantum states under semidefinite constraints is a pivotal problem located at the mathematical core of various applications in quantum information theory. An efficient method for providing  provable upper and lower bounds is the central result of this work.    
Our primordial motivation stems from the essential task of estimating secret key rates for QKD from the measurement statistics of a real device. Further applications  include the computation of channel capacities, the estimation of entanglement measures and many more. 
We build on a recently introduced integral representation of quantum relative entropy by [Frenkel, \textit{Quantum} 7, 1102 (2023)] and provide reliable bounds as a sequence  of semidefinite programs (SDPs). Our approach ensures provable sublinear convergence in the discretization, while also maintaining resource efficiency in terms of SDP matrix dimensions. Additionally, we can provide gap estimates to the optimum at each iteration stage.

\end{abstract}

\maketitle

\section{Introduction}
Within the last four decades the field of quantum cryptography has undertaken a massive evolution. Originating from theoretical considerations by Bennet and Brassard in 1984 \cite{Bennett2014} we are now in a world where technologies like QKD systems and Quantum random number generators are on the edge of being a marked ready reality. Moreover, there is an ongoing flow (see e.g. \cite{Nadlinger_2022,Zhang_2022} and references therein) of demonstrator setups and proof-of-principle experiments within the academic realm that bears a cornucopia of cryptographic quantum  technologies that may reach a next stage in a not too far future. 

Despite these gigantic leaps on the technological side, we have to constitute that the theoretical security analysis of quantum cryptographic systems is still in a process of catching up with these developments. To the best of our knowledge, there are yet no commercial devices with a fully comprehensive, openly accessible, and by the community verified security proof. 
Nevertheless, theory research has taken the essential steps in providing the building blocks for a framework that allows to do this \cite{RENNER2008}. Most notably, the development of the entropy accumulation theorem \cite{Dupuis2020,Metger2022} and comparable techniques \cite{Christandl2009}, allow us to deduce reliable guarantees on an $\varepsilon$-secure extractable finite key in the context of general quantum attacks requiring only bounds on an asymptotic quantity such as the conditional von Neumann entropy as input.  

The pivotal problem, and the input to this framework,  is to find a good lower bound  on the securely extractable randomness that a cryptographic device offers in the presence of a fully quantum attacker \cite{Senno_2023}. 
Mathematically this quantity is expressed by the conditional von Neumann entropy $H(X|E)$. Using Claude Shannon's intuitive description, it can be understood as the \textit{uncertainty} an attacker $E$ has about the outcome of a measurement $X$, which is performed by the user of a device. 
There are several  existing numerical techniques for estimating this quantity given a set of measurement data provided by a device \cite{Fawzi_2018,Hu_2022,Brown2024,Ara_jo_2023,Tan_2021,qics}. We will add to this collection, by providing a practical and resource efficient method for this problem, which interpolates between an executable tool and theoretical bounds on the relative entropy by convex interpolation.

At the core of our work stands a recently described \cite{Jencova2024,Frenkel2023}, and pleasingly elegant, integral representation
of the quantum (Umegaki) relative entropy \cite{Hiai1991} (see also \cite{Hirche2024}) that we employ in order to formulate the problem of reliably bounding $H(X|E)$ as an instance of semidefinite programs (SDP) by discretizing integrals. 
Our method comes with a provable sublinear convergence guarantee in the discretization, whilst staying resource efficient with the matrix dimension of the underlying SDPs. We furthermore can provide an estimate for the gap to the optimum for any discretization stage. 

To this end, let $\mathcal{H}\cong\mathbb{C}^d$ be a finite-dimensional Hilbert space. 
Write $\mathcal{B}(\mathcal{H})$ for the (bounded) linear operators on $\mathcal{H}$ and
$\mathcal{S}(\mathcal{H})\coloneqq\{\omega\in\mathcal{B}(\mathcal{H}) : \omega\ge 0,\ \tr[\omega]=1\}$
for the set of quantum states (density operators). Let $h_i:\mathcal{B}(\mathcal{H})\times\mathcal{B}(\mathcal{H})\to\mathbb{R}$ be affine maps, for $i=1,\dots,n$.
The central mathematical problem considered here—more general than estimating a conditional entropy $H(X|E)$ and not limited to QKD—is:
\begin{equation}\label{eq:general_problem}
\begin{aligned}
\inf \   & D(\rho\Vert\sigma) \\
\operatorname{s.t.}\ & h_i(\rho,\sigma)\ge 0,\qquad i=1,\dots,n,\\
                       & \mu\,\sigma \le \rho \le \lambda\,\sigma,\\
                       & \rho,\sigma\in\mathcal{S}(\mathcal{H}),
\end{aligned}
\end{equation}
where the (Umegaki) quantum relative entropy is $D(\rho\Vert\sigma)\coloneqq \tr\!\bigl[\rho\bigl(\log\rho-\log\sigma\bigr)\bigr]$. The constraint $\rho\le \lambda\,\sigma$ (with finite $\lambda$) enforces $\operatorname{supp}(\rho)\subseteq \operatorname{supp}(\sigma)$, ensuring the relative entropy to be finite; if, in addition, $\mu>0$, then $\operatorname{supp}(\rho)=\operatorname{supp}(\sigma)$.

Despite being convex, this optimisation problem is highly non-linear and contains the analytically benign, but numerically problematic matrix logarithm. Thus, for general instances, \eqref{eq:general_problem} can not be solved  directly by existing standard methods. The construction of a converging sequence of reliable lower bounds on the value $c$ in \eqref{eq:general_problem} is the central technical contribution of this work. 

Our focus task of estimating key-rates can be cast as an instance of this (see the last section of \autoref{sec:results} and supplementary material \autoref{apendix:sdp_formulations}). Here lower bounds  on \eqref{eq:general_problem} directly translate into lower bounds on the key-rate, which is exactly the direction of an estimate needed for a reliable security proof.  
There is however a long list of further problems that can be formulated as an instance of \eqref{eq:general_problem}. It includes for example the optimisation over all types of entropies which are expressible as relative entropies. For example we provide the calculation of the entanglement-assisted classical capacity of a quantum channel in the supplementary material \autoref{sec:quantum_shannon} where one has to optimise in fact the mutual information of a bipartite system.  
The optimization problem \eqref{eq:general_problem} naturally generalizes from relative entropies to general f-divergences. With minimal adjustments, our method can also tackle this  class.  Despite not be the focus task of this work,  a details numerical analysis is left for future work, we already formulated the relevant technical parts  of the Methods section \ref{sec:methods}  from this more general perspective. 

\section{Results}\label{sec:results}
In the following we denote by $\mathcal{B}(\mathcal{H})$ the set of (bounded) linear operators on a finite-dimensional Hilbert space $\mathcal{H}$ and $\mathcal{S}(\mathcal{H})$ the set of quantum states on $\mathcal{H}$, i.e. all positive operators with unit trace. The trace on $\mathcal{B}(\mathcal{H})$ is denoted as $\operatorname{tr}[\cdot]$. Moreover, any self adjoint operator $A \in \mathcal{B}(\mathcal{H})$, can be uniquely decomposed as a difference $A=A^+-A^-$ of Hilbert-Schmidt orthogonal positive operators $A^+$ and $A^-$. Let $\tr^+[A]\coloneqq\tr[A^+]$ denote the trace of the positive part of $A$ (similarly $\tr^-[A]\coloneqq  \tr[A^-] = \tr^+[-A]$). Note that this is an SDP given by
\begin{equation}\label{eq:variational_tr_plus}
\begin{aligned}
    \operatorname{tr}^+[A] = \sup \ &\tr[PA] \\
    \operatorname{s.t.} \ &0\leq P \leq \mathds{1}.
\end{aligned}
\end{equation}
In the following we make use of the representation    
\begin{align}\label{eq:jencova_integral}
    D(\rho \Vert \sigma) = \int_\mu^\lambda \frac{ds}{s} \tr^+[\sigma s - \rho] + \log \lambda + 1 - \lambda
\end{align}
which was firstly described by Jenčová in \cite{Jencova2024} and holds for pairs of quantum states that fulfil $\mu \sigma \leq \rho \leq \lambda \sigma$ with constants $\lambda>\mu \geq 0$.\footnote{Observe that we always use $\tr^+[\cdot]$ in comparison to $\tr^-[\cdot]$ in \cite{Jencova2024}. The reason for that is the SDP characterization in \eqref{eq:variational_tr_plus}, which can be written without a sign.} As outlined in the following, and with more detail in the methods section, the representation \eqref{eq:jencova_integral} can be used to reformulate the non-linear function $D(\rho\Vert\sigma)$ as solution to a semidefinite minimisation. The leading idea of our method is  then to incorporate this into \eqref{eq:general_problem} in order to obtain a SDP formulation of the whole problem. Along this path we make use of a discretisation of the integral in \eqref{eq:jencova_integral}. 
This discretisation introduces a set of free variational parameters into our method, and a suboptimal choice of these will produce a gap. 
This gap can however be quantified and the discretisation parameters can be adjusted iteratively leading to an increasing sequence of estimates on 
\eqref{eq:general_problem}.

\textit{-- Discretisation and SDP formulation:}\;
For an interval $(a,b)$ with $\mu<a<b<\lambda$ we have (see the discussion around \autoref{lem:properties_g}) the basic estimate
\begin{align}\label{eq:interval}
    \int_a^b \frac{ds}{s} \tr^+[\sigma s - \rho] \geq \tr^+[\sigma (b-a) + \rho\log(a/b)].
\end{align}
Based on \eqref{eq:interval},  we discretize the integral \eqref{eq:jencova_integral} on a grid  of points $\mathbf{t}=(t_1,\dots,t_r)$, i.e. intervals $(t_k,t_{k+1})$, and obtain an estimate on the relative entropy from below. We furthermore use that the evaluation of the functional $\tr^+[\cdot]$ can be formulated as an SDP, which in combination leads us to the following proposition:

\begin{prop}\label{prop:mainprop}
    For any grid  $\mathbf{t}$, with  $\mu \leq t_1 \leq \cdots \leq \cdots t_r = \lambda$, the relative entropy is bounded from below by the semidefinite optimisation
    \begin{equation}\label{eq:relax_sdp}
    \begin{aligned}
        D(\rho \Vert \sigma) \geq \inf \ &\sum_{k = 1}^{r-1} \tr[\mu_k] + \log \lambda + 1 - \lambda \\ 
         \operatorname{s.t.} \ & \mu_k \geq \alpha_k \rho + \beta_k \sigma,  \\ 
                     & \quad \quad \quad \quad \quad  k = 1, \dots, r-1 \\
                     &\mu_k\geq 0,
    \end{aligned}
    \end{equation}
     with coefficients 
     \begin{align}\label{eq:def_alphak_beta_k}
         \alpha_k= \log\left(\frac{t_k}{t_{k+1}}\right) \text{ and } \beta_k=t_{k+1}-t_k.
     \end{align}
\end{prop}
\begin{proof}
   Supplementary material: \autoref{appendix_proof_propmain}.
\end{proof}
\textit{-- Approximation of  \eqref{eq:general_problem}:}\;
We are now in a position to state the main mechanism of our method. Fixing $\mu,\lambda \in \mathbb{R}_{\geq 0}$\footnote{The knowledge of $\mu,\lambda \in \mathbb{R}_{\geq 0}$ guarantees that if the optimization problem \eqref{eq:general_problem} is feasible, it has already finite value. The value of $\lambda$ is a bound on the $D_{\operatorname{max}}$-relative entropy of the set of feasible states. In many applications this is known beforehand, e.g. in the key rate estimation it is given by Hayashi's pinching inequality.}, a grid $\mathbf{t}$ and combining \eqref{eq:relax_sdp} with \eqref{eq:general_problem} yields the SDP
\begin{align}\label{eq:relaxation_lower_bound}
    c_l\left(\mathbf{t}\right) \coloneqq \inf \  &\sum_{k = 1}^{r-1} \tr[\mu_k] + \log \lambda + 1 - \lambda\\ \nonumber
    \operatorname{s.t.} 
    \ & h_i(\rho,\sigma)\geq 0, \quad i=1,\dots,n \\ \nonumber
    \ &\mu_k \geq \alpha_k \rho + \beta_k \sigma,  \quad  k = 1, \dots, r-1\\ \nonumber
    \ &\mu \sigma \leq \rho \leq \lambda \sigma \\ \nonumber
    \ &\sigma,\rho \in \mathcal{S}(\mathcal{H}), \quad \mu_k\geq 0,
\end{align}
which is a lower bound on $c$ from \eqref{eq:general_problem}. Moreover, optimising over all grids $\mathbf{t}$ gives the tight bound
\begin{align}\label{eq:mainestimate-1}
    c=\sup_{\mathbf t\subset [\mu,\lambda]} c_l\left(\mathbf t\right). 
\end{align}
This reduces the task of approximating $c$ to the quest for a good grid $\mathbf t$. As every grid gives a valid lower bound by \autoref{prop:mainprop}, we are now freed to employ heuristic methods and still obtain rigorous statements, for example in a security proof. 

\textit{-- Upper bounds and a gap estimate:}\;
In order to construct an algorithm that terminates in finite time, it is helpful to give an estimate on the accuracy of an approximation. 
Similar to \autoref{prop:mainprop}, we can also construct semidefinite upper bounds for~$c$ (see supplementary material, \autoref{appendix_upper_bound}), now involving coefficients $\gamma_k, \delta_k \in \mathbb{R}$ as described in~\eqref{eq:coeff_upper}. Similarly to \eqref{eq:relaxation_lower_bound} we have
\begin{equation}\label{eq:relaxation_upper_bound}
\begin{aligned}
   c_u\left(\mathbf t \right) \coloneqq \inf \ &\sum_{k = 1}^{r} \tr[\nu_k] + \log \lambda + 1 - \lambda\\ 
   \operatorname{s.t.}
    \ & h_i(\rho,\sigma)\geq 0 \quad i=1,\dots,n \\ 
    \ &   \nu_k \geq \gamma_k \rho + \delta_k \sigma,  \quad k = 1,\dots, r. \\ 
    \ &\mu \sigma \leq \rho \leq \lambda \sigma \\ 
    \ &\sigma,\rho \in \mathcal{S}(\mathcal{H}), \quad \nu_k\geq 0.
\end{aligned}
\end{equation}
Concluding \eqref{eq:general_problem}, \eqref{eq:relaxation_upper_bound} and \eqref{eq:relaxation_lower_bound}, we get the chain of inequalities $c_l \left(\mathbf t \right) \leq c \leq c_u \left(\mathbf t \right)$ and a gap estimator 
\begin{align}\label{eq:gap}
    \Delta\left(\mathbf t \right)= c_u\left(\mathbf t \right)-c_l\left(\mathbf t \right).
\end{align}

\textit{-- Simple methods with convergence guarantee :}\;

As our methods for establishing the lower bound~\eqref{eq:relaxation_lower_bound} and the upper bound~\eqref{eq:relaxation_upper_bound} rely on estimates of an integral, convergence can be guaranteed if we are able to provide uniform bounds on the integrand for all pairs of feasible states $\rho,\sigma \in \mathcal{S}(\mathcal{H})$.\footnote{Again, we observe that the bound $\lambda$ on $D_{\operatorname{max}}(\rho \Vert \sigma)$ is essential; without it, states with orthogonal supports would immediately preclude such uniformity.} This is precisely why the integral representation from~\cite{Jencova2024} is particularly useful: it yields a compact integration interval $[\mu,\lambda]$, provided we can bound the $\operatorname{D}_{\operatorname{max}}$-relative entropy for an optimal pair of states $\rho^\star$ and $\sigma^\star$ solving~\eqref{eq:general_problem}. 

The final missing ingredients, beyond compactness, to guarantee uniform convergence in our setting are provided in \autoref{lem:properties_g}, where we show that $g(s) \coloneqq \operatorname{tr}^+[\sigma s - \rho]$ is convex, monotonically increasing, and Lipschitz continuous. With these tools in place, we prove in \autoref{prop:convergence_f_div} that the upper bounds converge uniformly, since they are nothing more than a convex interpolation of the function $g(s)$. 

The outcome of this discussion of tools is summarized in the following corollary.

\begin{corr}\label{cor:convergence_xlogx}
Let $\rho,\sigma\in\mathcal{S}(\mathcal{H})$ with $\mu\sigma\le\rho\le\lambda\sigma$ and $\mu>0$\footnote{Actually choosing $\mu = 0$ yields nothing special, because then we can apply our method on the interval $[\varepsilon,\lambda]$ and get an error $\varepsilon$ in the approximation on the interval $[0,\varepsilon]$. For details see \autoref{cor:qkd}.}.
Choose the grid recursively by\footnote{For an explicit explanation how the grid construction works in practice, see the section "Applications to QKD".}
\begin{equation}\label{eq:def_special_grid_xlogx}
t_k=
\begin{cases}
\mu, & k=1,\\
t_{k-1}+\sqrt{8\varepsilon\,t_{k-1}}, & k\ge 2.
\end{cases}
\end{equation}
Then the total approximation error obeys
\begin{align}
c_u(\mathbf{t}) - c \le \varepsilon,
\end{align}
and the number of grid points satisfies the explicit bound
\begin{align}
r = \mathcal{O}\bigg(\sqrt{\frac{\lambda}{\varepsilon}}\bigg).
\end{align}
\end{corr}

\begin{proof}
Using \autoref{prop:convergence_f_div} with $f''(s)=1/s$ yields convergence of the interpolation: 
use $\int_{\mu}^{\lambda}\sqrt{f''(s)}\,ds=\int_{\mu}^{\lambda}s^{-1/2}ds
=2(\sqrt{\lambda}-\sqrt{\mu})$, and since $1/s$ is decreasing on $[\mu,\lambda]$,
$L_{k-1}=1/t_{k-1}$, yielding \eqref{eq:def_special_grid_xlogx}. Moreover, \autoref{cor:value_convergence} yields that uniform convergence of the bounds is enough in order to prove that the values of \eqref{eq:relaxation_upper_bound} converge to $c$ in \eqref{eq:general_problem}.
\end{proof}

\begin{figure*}
    \centering
        \includegraphics[width=0.8\linewidth]{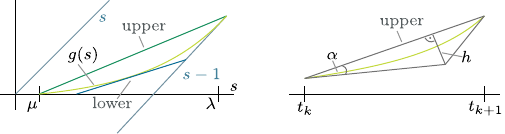}
    \caption{\justifying The left picture shows that we can approximate a monotone and convex function from below with linear functions. It furthermore shows the corridor in which the divergence will be located. Furthermore, there is a degree of freedom in choosing a tangential straight from below. A mirrored straight $g^{\prime \prime}$, which is a feasible lower bound yields the same convergence rate for the lower bound.
    The worse case that could happen for approximation is that the function has a kink as shown on the right picture. The right picture shows an interval $[t_k,t_{k+1}]$. Then we see the error for the upper bound scales with the volume of a blunt triangle.}\label{fig:convergence}
\end{figure*}

Moreover, as discussed rigorously in \autoref{sec:methods}, the lower bounds can be interpreted as a specific type of optimal \emph{supporting lines} (see \autoref{sec:methods} for a precise definition), i.e., tangents lying below $g(s)$ with respect to an integral norm determined by the weight function $s \mapsto \frac{1}{s}$. Geometrically, this can be visualized via a mirroring argument in \autoref{fig:convergence}. Consequently, the estimate 
\begin{align}
c - c_l(\mathbf{t}) \le \varepsilon
\end{align}
follows immediately from \autoref{cor:convergence_xlogx}.

We can conclude this section with the result that in a numerical algorithm the gap in \eqref{eq:gap} is in the magnitude of $\varepsilon$ if we choose $\mathcal{O}\bigg(\sqrt{\frac{\lambda}{\varepsilon}}\bigg)$ many grid points in the discretization.

\textit{--  Heuristic methods:}\;
Motivated by the observation that our approximation reduces to a linear program when $\rho$ and $\sigma$ commute—and, in particular, $\tr^+[\sigma s - \rho]$ is then piecewise linear, i.e., a sum of affine segments combined via a pointwise maximum—we conclude that in this case a finite set of grid points already suffices for an exact result, due to the affine nature of our approximations. Consequently, a heuristic should also allow for routines that drop points from $\mathbf{t}$ in order to remain resource-efficient. 

This is especially relevant for the inner approximation, i.e., the upper bounds. In fact, one can delete all grid points except the one corresponding to the current optimizer from the previous iteration, since the upper bound is a continuous function of $s$. This yields a highly efficient heuristic for obtaining good upper bounds, made possible by the fact that we approximate the curves defined by $\tr^+[\sigma s - \rho]$ from above using a convex, continuous function. In comparison to the upper bounds, the lower bound \eqref{eq:relaxation_lower_bound} is not continuous. For this reason it is impossible to delete grid points. Therefore it becomes even more important to control the grid points wisely. An additional, but not rigorous way of getting the sequence of values monotone is that we can include a convex constraint such that the solver is enforced to stay monotone. Of course this destroys the fact that we want provable upper or lower bounds. But interestingly one can enforce monotony for a couple of rounds, then using the resulting pair of optimal states as a warm start without this constraint. This method is efficient and leads to good results\footnote{To the best of our knowledge, warm starts with cvx are not possible in MATLAB. However, in Python it is possible and we run this heuristic in a Python program. }. 

\begin{figure*}
    \centering
    \subfloat{
            \begin{tikzpicture}
\begin{axis}[width=6cm,height=6cm,scatter/classes={%
    a={mark=o}},
    ymode = log,
    xmode = log,
    xlabel = grid points,
    ylabel = error]
\addplot[only marks,%
    scatter src=explicit symbolic,color=PineGreen]%
table{results/inner_list.txt};
\addlegendentry{upper bound}
\addplot[only marks,%
    scatter src=explicit symbolic,color=MidnightBlue]%
table{results/outer_list.txt};
\addlegendentry{lower bound}
\addplot[smooth,color=black]%
table{results/regression.txt};
\addlegendentry{regression}
\end{axis}
\end{tikzpicture}
    }      
    \subfloat{
           \begin{tikzpicture}
\begin{axis}[width=6cm,height=6cm,scatter/classes={%
    a={mark=o}},
    xlabel = iterations,
    ylabel = relative entropy]
\addplot[only marks,%
    scatter src=explicit symbolic,color=PineGreen]%
table{results/inner_out.txt};
\addlegendentry{upper bound}
\addplot[only marks,%
    scatter src=explicit symbolic,color=MidnightBlue]%
table{results/outer_out.txt};
\addlegendentry{lower bound}
\end{axis}
\end{tikzpicture}
    }
      \subfloat{
            \begin{tikzpicture}
            
\begin{axis}[width=6cm,height=6cm,scatter/classes={%
    a={mark=o}},
  xlabel=alpha,
  ylabel=relative entropy,
  legend style={draw=none}]
\addlegendentry{local dimension $2$}
\addplot[scatter,only marks,%
    scatter src=explicit symbolic,
   mark options={YellowGreen},
  error bars/.cd, 
    y fixed,
    y dir=both, 
    y explicit
]
table[x=x,y=y,y error = error]{lower_bounds_device_dependent/2qubits/dim2_fin.txt};
\addplot[scatter,only marks,%
    scatter src=explicit symbolic,
   mark options={SeaGreen},
  error bars/.cd, 
    y fixed,
    y dir=both, 
    y explicit
]
table[x=x,y=y,y error = error]{lower_bounds_device_dependent/4qubits/dim4_fin.txt};
\addlegendentry{local dimension $4$}
\addplot[scatter,only marks,%
    scatter src=explicit symbolic,
   mark options={BlueGreen},
  error bars/.cd, 
    y fixed,
    y dir=both, 
    y explicit]
table[x=x,y=y,y error = error]{lower_bounds_device_dependent/6qubits/dim8_fin.txt};
\addlegendentry{local dimension $8$}
\end{axis}
\end{tikzpicture}

    }
\caption{ \justifying The left plot shows the sublinear convergence rate in the discretization for a fixed state pair $\rho$ and $\sigma$ with the method from \autoref{cor:convergence_xlogx} in dimension $4$. To be concrete, we calculated the relative entropy by its definition and the error between our approximation and this value for increasing number of grid points resulting from the iterative formula in \autoref{cor:convergence_xlogx}.
In the middle plot, we show a generic instance of~\eqref{eq:general_problem} as discussed in~\eqref{eq:witness_instance}. All numerical examples are available in the \href{https://github.com/gereonkn/relative-entropy-optimization.git}{GitHub repository}. 
The right plot shows the extractable randomness from~\eqref{eq:relaxation_lower_bound_qkd} as a function of the parameter~$\alpha$ for a state defined in~\eqref{eq:def_alpha_state} and a pair of mutually unbiased bases for Alice and Bob in various local dimensions up to~$8$, corresponding to a total dimension of~$64$. The plot exhibits the expected behaviour with respect to depolarizing noise, parameterized by $\alpha \in [0,1]$.}
\label{fig:results}
\end{figure*}

\textit{-- Application to Quantum Key Distribution :}\;

\begin{figure}
    \centering
\begin{tikzpicture}
\begin{axis}[
    ybar,
    bar width=10pt,
    enlarge x limits=0.20,
    symbolic x coords={d=2,d=4,d=8},
    xtick=data,
    ymin=0,
    ylabel={Runtime (s)},
    title={QKD Runtime Estimate},
    legend pos=north west,
    legend cell align={left},
    axis line style={black!70},
    tick style={black!70},
]
\addplot+[ybar, draw=MidnightBlue!80, fill=MidnightBlue!70] 
  coordinates {(d=2,1.494754) (d=4,12.915239) (d=8,209.147548)};
\addplot+[ybar, draw=PineGreen!80, fill=PineGreen!70] 
  coordinates {(d=2,3.231543) (d=4,48.435805) (d=8,nan)};
\addplot+[ybar, draw=DeepTeal!80, fill=DeepTeal!70] 
  coordinates {(d=2,1.6301770210266113) (d=4,1.9859693050384521) (d=8,31.78383755683899)};

\legend{Our Work, Araujo, QICS}
\end{axis}
\end{tikzpicture}

\vspace{1cm}

\begin{tikzpicture}
\begin{axis}[
    ybar,
    bar width=10pt,
    enlarge x limits=0.20,
    symbolic x coords={d=2,d=4,d=8},
    xtick=data,
    ymin=0,
    ylabel={Precision ($-\log_{10}(\log_2(d)-\operatorname{outcome)}$)},
    title={QKD Quality of Estimates},
    legend pos=north west,
    legend cell align={left},
    axis line style={black!70},
    tick style={black!70},
]

\addplot+[ybar, draw=MidnightBlue!80, fill=MidnightBlue!70] 
  coordinates {(d=2,4.1419) (d=4,3.9658) (d=8,0.8988)};

\addplot+[ybar, draw=PineGreen!80, fill=PineGreen!70] 
  coordinates {(d=2,3.7565297459) (d=4,4.1355357738) (d=8,nan)};

\addplot+[ybar, draw=DeepTeal!80, fill=DeepTeal!70] 
  coordinates {(d=2,7.616885779545696) (d=4,7.427209434815066) (d=8,7.708803652051622)};

\end{axis}
\end{tikzpicture}
    \caption{We compare the key rate protocol for entanglement based QKD for local dimensions $2,4$ and $8$ with the techniques from \cite{Ara_jo_2023}, our techniques and \cite{qics}. The first plot shows runtime estimates between all of the three methods and the second plots shows the precision in logarithmic scale. The system is equipped with a 13th Gen Intel\textsuperscript{\textregistered} Core\texttrademark{} i3-1315U processor and 8\,GB of RAM. The metod by \cite{Ara_jo_2023} was not executable for local dimension $4$ and $8$, such that we replaced the values for $4$ with values for local dimension $3$. 
}
    \label{fig:comparison_methods}
\end{figure}
The instances that initially motivate us to investigate \eqref{eq:general_problem} arise from the task of estimating the extractable randomness for applications in quantum cryptography. 
Consider a system consisting of three Hilbert spaces  $\mathcal{H}_{ABE}\coloneqq\mathcal{H}_A \otimes \mathcal{H}_B \otimes \mathcal{H}_E$.
In a basic entanglement-based QKD-setting two parties, say, Alice and Bob, perform measurements $X^A_0,\dots, X^A_n $ and $X^B_0,\dots, X^B_n $ on their shares of a tripartite quantum state $\psi_{ABE}\in \mathcal{H}_{ABE}$ provided by a third malicious party Eve.  Following common conventions, the outcomes of measurements $X^A_0 X^B_0$ will be used to generate a key, whereas the data from all other measurements is used to test properties of the state $\psi_{ABE}$, and by this, bound the influence of Eve.  
For error correction, it is assumed that Alice's data, i.e. the outcomes of $X^A_0$, correspond to the correct key, which means that Bob has to correct the data arising from the measurement $X^B_0$. 
Furthermore, we will employ that each measurement $X^S_i$ can be modelled by a channel $\Phi^S_{i}:\mathcal{S}(\mathcal{H}_S)\to \mathcal{S}(\mathcal{R}_{X^S_i})$ that maps states from a quantum system $\mathcal{H}_S$ to a probability distribution $p_{S,i}$ on a classical register $\mathcal{R}_{X^S_i}$. See also refs \cite{Tan_2021,Winick_2018} for more details on this model. 

Within the notation above, the securely extractable randomness of Alice's key measurement is given by the conditional entropy 
$H(X^A_0|E)_{(\Phi^A_{0}\otimes id_E )[\rho_{AE}]}$ and depends on the reduced quantum state $\rho_{AE}$ of the Alice-Eve system. Lower bounds on this quantity, which is up to now only defined in an asymptotic scenario, are essential for reliably bounding key rates in a full QKD setting involving multiple rounds. This accounts for the asymptotic regime, in which the Devetak-Winter formula \cite{Devetak_2005} can be used, as well as for finitely many rounds under collective attacks, where the AEP can be used \cite{Tomamichel_2009}, and general attacks where either EAT \cite{Dupuis2020,Metger2022} or de Finetti based methods can be employed \cite{Renner_Cirac_2009,Christandl2009,ArnonFriedman2015}.   

Using a technical result for calculating the entropy of a state $\rho_{AE}$ \cite{Tomamichel_2016,Tan_2021,Winick_2018} (see \autoref{lem:Lower_bounds}), we can express the conditional von Neumann entropy in terms of a relative entropy  
\begin{align}\label{eq:without_eve}
    H(X^A_0|E)_{\Phi^A_{0} [\rho_{AE}]}
    = D(\rho_{AB}\Vert  \Phi^A_{0} [\rho_{AB}]).
\end{align}

Test data obtained from additional measurements $X^S_i$ naturally give affine constraints on an unknown state $\rho_{AB}$. 
The central problem of lower bounding the extractable randomness can therefore be formulated as   
\begin{equation}\label{eq:crypto_problem}
\begin{aligned}
    \inf \ &D(\rho_{AB} \Vert \Phi^A_{0}[\rho_{AB}] )  \\ 
    \operatorname{s.t.} \  & \Phi^A_{i} \otimes \Phi^B_{j} [\rho_{AB}] = p_{AB,ij} \quad  i,j =  1,\dots,n  \\ 
    &\rho_{AB} \in \mathcal{S}(\mathcal{H}_A \otimes \mathcal{H}_B) \\ 
    &0 \leq \rho_{AB} \leq \lambda \Phi^A_{0}[\rho_{AB}]
\end{aligned}
\end{equation}
which is an instance of \eqref{eq:general_problem} to which we can apply our methodology. In order to solve \eqref{eq:crypto_problem}, we need to fix constants $\mu$ and $\lambda$ such that an optimal minimizer $\rho_{AB} \in \mathcal{S}(\mathcal{H}_A \otimes \mathcal{H}_B)$ satisfies
\begin{align}\label{eq:inequality_lambda_mu_qkd}
    \mu \Phi^A_0(\rho_{AB}) \leq \rho_{AB} \leq \lambda \Phi^A_0(\rho_{AB}).
\end{align}
By Hayashi's pinching inequality, the value of $\lambda$ can be chosen as the square root of the overall dimension and is therefore known in advance. Without loss of generality, the value of $\mu$ in~\eqref{eq:inequality_lambda_mu_qkd} can be set to zero. In the numerical examples in the \href{https://github.com/gereonkn/relative-entropy-optimization.git}{repository}, the function \texttt{grid\_function(c, epsilon, mu, lamb)} generates a sequence of grid points starting from the lower bound $\mu$ and ending at the upper bound $\lambda$ as defined in \autoref{cor:convergence_xlogx}. At each step, the next grid point is computed as
\begin{align}
    t_{k} = t_{k-1} + \sqrt{\frac{t_{k-1} \, \epsilon}{c}},
\end{align}
where $t_k$ denotes the current grid point. This process continues until the next point would exceed $\lambda$, at which stage the final grid value is set exactly to $\lambda$. The resulting sequence is returned as the vector \texttt{grid}.  

With this in hand, we can formulate the following explicit optimization program (with $\alpha_k,\beta_k \in \mathbb{R}$ computed as in \eqref{eq:def_alphak_beta_k}) for provable lower bounds on \eqref{eq:crypto_problem}:
\begin{align}\label{eq:relaxation_lower_bound_qkd}
    \inf \  &\sum_{k = 1}^{r-1} \tr[\mu_k] + \log \lambda + 1 - \lambda\\ \nonumber
    \operatorname{s.t.} 
    \ & \Phi^A_{i} \otimes \Phi^B_{j} [\rho_{AB}] = p_{AB,ij} \quad  i,j = 1,\dots,n, \\ \nonumber
    \ &\mu_k \geq \alpha_k \rho_{AB} + \beta_k \Phi_A^0(\rho_{AB})\quad k = 1,\dots, r-1, \\ \nonumber
    \ &\rho_{AB} \in \mathcal{S}( \mathcal{H}_{A}\otimes \mathcal{H}_B), \quad \mu_k\geq 0.
\end{align}
To construct a probability distribution $p_{AB,ij}$ as test data in the  following examples, we start from a maximally entangled state $\Omega_{AB} \in \mathcal{S}( \mathcal{H}_{A}\otimes \mathcal{H}_B)$ and mix it with white noise, obtaining
\begin{align}\label{eq:def_alpha_state}
    \Omega_{AB}(\alpha) \coloneqq (1-\alpha)\,\Omega_{AB} + \alpha\,\frac{\mathds{1}}{d}.
\end{align}
As measurement channels, we employ projective measurements in two mutually unbiased bases (see e.g. \cite{Klappenecker2004}). Solutions of \eqref{eq:relaxation_lower_bound_qkd} are shown in \autoref{fig:results} the right plot for local dimensions, i.e. dimensions of Alice respectively Bob's system between $2,4,8$, which corresponds to $1,2$ and $3$ qubits per party and different values for $\alpha \in [0,1]$. All results are achieved in seconds on a personal computer. 
Moreover, we can state a similar corollary as \autoref{cor:convergence_xlogx} for the special case of QKD:
\begin{corr}\label{cor:qkd}
Let $\rho\in\mathcal{S}(\mathcal{H}_A \otimes \mathcal{H}_B)$ and $\Phi_0^A$ a projective measurement channel with $d_A = \dim \mathcal{H}_A$ many outcomes and choose $\varepsilon>0$ fixed. Furthermore, choose the grid recursively by
\begin{equation}
t_k=
\begin{cases}
\varepsilon, & k=1,\\
t_{k-1}+\sqrt{8\varepsilon\,t_{k-1}}, & k\ge 2.
\end{cases}
\end{equation}
Then the total approximation error for \eqref{eq:relaxation_lower_bound_qkd} obeys
\begin{align}\label{eq:bound_qkd}
c-c_l(\mathbf{t}) \le 2 \varepsilon,
\end{align}
and the number of grid points satisfies the explicit bound
\begin{align}
r = \mathcal{O}\bigg(\sqrt{\frac{d_A}{\varepsilon}}\bigg).
\end{align}
\end{corr}
\begin{proof}
    From Hayashi's pinching inequality \eqref{eq:inequality_lambda_mu_qkd}, it follows immediately that we may choose $\lambda = d_A$.  
Moreover, on the interval $[\varepsilon, \lambda]$, we apply \autoref{cor:convergence_xlogx} to obtain an approximation error $\varepsilon$. 
Additionally, using the estimate
\begin{equation}
\begin{aligned}
    \operatorname{tr}^+\big[\Phi_0^A (\rho_{AB})\, s - \rho_{AB}\big]  
        &= \sup_{0\leq P \leq \mathds{1}} \operatorname{tr}\!\big[P\big(\Phi_0^A (\rho_{AB})\, s - \rho_{AB}\big)\big] \\
        &= \sup_{0\leq P \leq \mathds{1}} \big( \operatorname{tr}[P\,\Phi_0^A (\rho_{AB})]\, s \\
        &\hspace{3cm}- \operatorname{tr}[P\,\rho_{AB}] \big) \\
        &\leq \sup_{0\leq P \leq \mathds{1}} \operatorname{tr}[P\,\Phi_0^A (\rho_{AB})]\, s \\
        &\leq s ,
\end{aligned}
\end{equation}
we can bound the contribution on the interval $[0, \varepsilon]$ as
\begin{align}
    \int_0^\varepsilon \frac{ds}{s} \, \operatorname{tr}^+\big[\Phi_0^A (\rho_{AB})\, s - \rho_{AB}\big]
    \leq \int_0^\varepsilon ds
    = \varepsilon .
\end{align}
This shows that the additional error incurred by assuming $\mu = 0$ is at most $\varepsilon$.  
Combining these bounds yields \eqref{eq:bound_qkd}.
\end{proof}

\textit{-- Further optimisation tasks that can be handled :}\;

As instances for the left and middle plots in \autoref{fig:results}, we use a randomly generated matrix $M$ (available in the \href{https://github.com/gereonkn/relative-entropy-optimization.git}{repository}) as a witness and solve problems of the form
\begin{align}\label{eq:witness_instance}
    c \coloneqq \inf \ & D(\rho \Vert \sigma)  \\ \nonumber
    \operatorname{s.t.} \ &\operatorname{tr}[\rho M] \geq \kappa_1,  \\ \nonumber
    &\operatorname{tr}[\sigma M] \leq \kappa_2,  \\ \nonumber
    &\mu \sigma \leq \rho \leq \lambda \sigma, \\ \nonumber
    &\sigma,\rho \in \mathcal{S}(\mathcal{H}), 
\end{align}
for various values of $\kappa_1, \kappa_2, \mu, \lambda \in \mathbb{R}$.  

The left plot shows the sublinear convergence in the discretization predicted by \autoref{cor:convergence_xlogx} in dimension~$4$. We plot the error corresponding to the grid from~\eqref{eq:def_special_grid_xlogx} as a function of the number of grid points for a generic instance. Furthermore, in the left plot of \autoref{fig:results}, we perform a regression with the model function $n \mapsto \frac{c}{n^2}$ for a regression parameter $c$. The analysis shows that $c$ is close to the chosen $\lambda$, as predicted by \autoref{cor:convergence_xlogx}. This supports the conclusion that we have obtained the correct asymptotic convergence behaviour.

Further instances of~\eqref{eq:general_problem} are reported in \autoref{apendix:sdp_formulations}. Notable examples include bounds on the relative entropy of entanglement, 
\begin{align}\label{eq:relative_entropy_of_entanglement_main_text}
    \min_{\sigma_{AB} \in \operatorname{SEP}(A:B)} D(\rho_{AB} \Vert \sigma_{AB}),
\end{align}
where $\operatorname{SEP}(A\!:\!B)$ denotes the set of separable states and $\rho_{AB}$ is a possibly entangled state, as well as the classical capacity of a quantum channel (see also~\cite{Fawzi_2018}).

\section{Discussion}\label{sec:discussion}

    The general optimisation problem in~\eqref{eq:general_problem} is a central task in (quantum) information theory, as it encompasses all types of numerical estimates for which the relative entropy is the underlying quantity. A well-known example is given by~\eqref{eq:crypto_problem}, where the optimisation yields bounds on the extractable randomness in a QKD experiment (or, similarly, in a random number generation experiment). Recently, instances of~\eqref{eq:general_problem} such as the relative entropy of entanglement in~\eqref{eq:relative_entropy_of_entanglement_main_text} and various channel-related quantities, including the classical capacity of a quantum channel, have also attracted attention. But, as already noted, the problem~\eqref{eq:general_problem} is a nonlinear yet convex optimisation problem. Its difficulty stems from the fact that the relative entropy is only lower semicontinuous and, for certain pairs of states $(\rho,\sigma) \in \mathcal{S}(\mathcal{H}) \times \mathcal{S}(\mathcal{H})$ with $D_{\operatorname{max}}(\rho \Vert \sigma) = \infty$, takes the value $+\infty$. This property complicates the use of generic convex optimisation solvers for tackling~\eqref{eq:general_problem}. 

    Given the importance of~\eqref{eq:general_problem}, a variety of solutions have been developed with different purposes and techniques. Broadly speaking, these approaches can be classified into four categories according to their scope and the tools they employ.

    The first class of methods includes the approach of~\cite{Fawzi_2018}, which estimates the relative entropy via the formula $D(\rho \Vert \sigma) = \operatorname{tr}[\rho (\log \rho - \log \sigma)]$ by numerically computing the matrix logarithm. This method is highly flexible but, for $\rho, \sigma \in \mathcal{S}(\mathcal{H})$ with $\dim \mathcal{H} = d$, it requires working with matrices of size $d^2 \times d^2$. The second class, aimed particularly at estimating~\eqref{eq:crypto_problem}, includes the works~\cite{Coles2016,Winick_2018,Hu_2022}. The most recent of these~\cite{Hu_2022} achieves very high precision (see Sec.~5, numerical testing, in~\cite{Hu_2022}) for systems consisting of one qubit per party (i.e., Alice’s and Bob’s systems), but is restricted to equality constraints in~\eqref{eq:crypto_problem}. Inequality constraints are mentioned as an outlook for future work. The third class is represented by the recent work~\cite{qics}, which introduces a solver based on self-concordant barriers for specialized cones associated with the quantum (relative) entropy. This framework enables the direct application of interior-point methods to~\eqref{eq:general_problem}. Finally, in the fourth class, \cite{Ara_jo_2023} and our present work use integral representations of the relative entropy. Specifically,~\cite{Ara_jo_2023} employs Kosaki’s formula~\cite{Kosaki}, while we use Frenkel’s formula~\cite{Frenkel2023}, approximating the relative entropy via numerical quadrature for the resulting integral representation. 

    With our technique, we propose, on the one hand, a concrete numerical tool for solving~\eqref{eq:general_problem}, which we benchmark against state-of-the-art instances in the QKD setting in~\autoref{fig:comparison_methods}, comparing with~\cite{qics} and~\cite{Ara_jo_2023}. On the other hand, we introduce a technical method to estimate the relative entropy variationally in both directions, providing both upper and lower bounds. 

    In the numerical benchmark, we observe that for small instances our technique achieves precision comparable to~\cite{Ara_jo_2023}, but is strictly outperformed by the specialized solver~\cite{qics} for relative entropy programming. In terms of runtime, our method shows a clear improvement over~\cite{Ara_jo_2023}, while still yielding slower performance than~\cite{qics}. 
    Beyond the numerical benchmark for~\eqref{eq:general_problem}, it is also instructive to compare our technique with~\cite{Ara_jo_2023} from a technical perspective, since both approaches rely on integral representations and thus offer comparable flexibility. In particular, our method provides provable upper \emph{and} lower bounds for~\eqref{eq:general_problem}, whereas~\cite{Ara_jo_2023} yields only upper bounds. This has the advantage that we can, for example, employ the dimension-efficient formula~\eqref{eq:without_eve} in the QKD setting, and we can quantify a gap as in~\eqref{eq:gap}. Moreover, the flexibility of our technique has enabled the first numerically applicable algorithm for obtaining provable bounds on the relative entropy of channels—an open problem since the development of resource theories for quantum channels \cite{koßmann2024semidefiniteoptimizationquantumrelative}.

    Moreover, both techniques enable applications to device-independent quantum key distribution, as reported in~\cite{Brown2024} and~\cite{koßmann2024boundingconditionalvonneumannentropy}, which is not possible with~\cite{qics}. In this regime \cite{koßmann2024boundingconditionalvonneumannentropy} reports an efficiency advantage of our technique. From the perspective of divergences and despite the coincidental overlap in terms of the Umegaki relative entropy, our tools and those of~\cite{Ara_jo_2023} are complementary in a broad sense: Kosaki’s integral formula can be extended to all operator monotone functions (see~\cite[Lem.~2.1]{Kosaki}), while our approach can be generalized to all $f$-divergences as shown in~\autoref{sec:methods}. 

    Concretely regarding our technique, the fact that we need fixed integral bounds $0<\mu\leq \lambda$ which on first view seems to be a disadvantage, turns out to be the important ingredient for a rigorous numerical analysis (i.e. theoretical error bounds). The existence of these values bounds the problem to finite range and one can think about the lower respective upper bounds as continuous functionals with values in a compact set. Therefore a rigorous numerical analysis becomes applicable. It is a beautiful observation that compactness of the image of the functionals is equivalent to finite relative entropy. Since we are only interested in minimisation tasks here, we get rid of numerical analysis artifacts with infinities directly and naturally.
    
    In contrast, controlling the number and places of supporting points is in general a difficult game with no a priori best solution. Of course one has to have in mind that practically the number of grid points must not be too big, because it increases the number of variables in the SDP solver directly. This calls for a clever heuristic, especially with regard to even larger dimensions. With the proofs of \autoref{prop:convergence_f_div} we give a clear mathematical, and therefore rigorous, framework which one can use in constructing heuristics. In the error analysis of \autoref{prop:convergence_f_div} we observe that it highly depends on $f^{\prime \prime}$ and its values on the grid intervals. For the function $s\mapsto s\log s$ the second derivative is given by $s \mapsto \frac{1}{s}$ and thus the error of our tools decrease as $s$ becomes larger. In particular, if $s>1$ the weight function $s \mapsto \frac{1}{s}$ in the integration is a damping factor. Thus, good heuristics for the Umegaki relative entropy should have a more refined grid for $s \in [\mu,1]$ and a coarser grid in $[1,\lambda]$. Moreover, one could ask for an optimal quadrature rule regarding this specific type of integrands $s \mapsto f^{\prime \prime}(s) \operatorname{tr}^+[\sigma s - \rho]$ for $f$ twice continuously-differentiable. 
    
    Another key could be to design a method that removes grid points as well. Many scenarios are possible here, which we leave open for future adjustment. In addition to heuristics, we would like to mention that our approach can also be carried out directly with the original integral representation of Frenkel \cite{Frenkel2023}. Since the singularities at $0$ and $1$ play a decisive role there, it becomes much more difficult to extract provable scenarios. However, we did numerical experiments in this direction with success, but apparently without numerical analysis, i.e. theoretical error-dependencies. 

    We conclude with an outlook for future research. In terms of grid refinement, we believe that further improvements are possible, and that more advanced numerical quadrature techniques for the integration step in \autoref{prop:convergence_f_div} and \autoref{cor:convergence_xlogx} may be applicable. From an information-theoretic perspective, a recent series of results has clarified how the family of $\alpha$-$f$-divergences defined in~\cite{Frenkel2023} relates to the well-known $\alpha$-relative entropies—namely, the Petz and sandwiched relative entropies (see~\cite{beigi2025propertiesapplicationsnewquantum,liu2025layercakerepresentationsquantum} for inequalities and~\cite[Thm.~3.2]{Hirche2024} for a regularization result). These findings suggest that our numerical techniques could be applied to estimate well-known $\alpha$-entropies. Developing this connection and applying it to finite-resource tasks in information theory would be an interesting direction for future work.

\section{Methods}\label{sec:methods}

As our tools easily  generalize to $f$-divergences, we provide here the fully detailed analysis for general $f$-divergences. A \emph{$f$-divergence} is defined as 
\begin{equation}\label{eq:def_f_divergence}
\begin{aligned}
    D_f(\rho \Vert \sigma) &\coloneqq \int_1^\infty f^{\prime \prime}(s) \operatorname{tr}^+[\rho - s\sigma] \\
    &\hspace{2cm} + s^{-3}f^{\prime \prime}(s^{-1})\operatorname{tr}^+[\sigma - s \rho] \ ds,
\end{aligned}
\end{equation} 
where $f : (0,\infty) \to \mathbb{R}$ is assumed to be twice continuously differentiable with $f(1) = 0$ (see, e.g., \cite[Def.~2.4]{Hirche2024}). From \cite[Prop.~2.6]{Hirche2024} it then follows that $f$-divergences are jointly convex in $\rho, \sigma \in \mathcal{S}(\mathcal{H})$, satisfy the data-processing inequality (DPI) for positive trace-nonincreasing linear maps, and are faithful. As usual, we define for states $\rho,\sigma\in \mathcal{S}(\mathcal{H})$
\begin{align}
    D_{\operatorname{max}}(\rho\Vert \sigma) \coloneqq \inf \{\lambda > 0 \ \vert \ \rho \leq e^\lambda \sigma\}
\end{align}
where we use the convention $\inf \{\} = \infty$.
We start with a first small result, generalizing \cite[Cor. 1]{Jencova2024} from the Umegaki relative entropy to general $f$-divergences.

\begin{prop}\label{prop:fd_general_formula}
For states $\rho,\sigma \in \mathcal{S}(\mathcal{H})$ define
$\mu \coloneqq e^{-D_{\max}(\sigma\Vert\rho)}$ and $\lambda \coloneqq e^{D_{\max}(\rho\Vert\sigma)}$.
Then we have
\begin{equation}\label{eq:fd_negpart}
\begin{aligned}
D_f(\rho\Vert\sigma)
&= \int_{\mu}^{\lambda} f^{\prime \prime}(s) \tr^+\big[s\sigma-\rho] \ ds \\
&\hspace{1cm}+ f(\lambda)+(1-\lambda)f^{\prime }(\lambda).
\end{aligned}
\end{equation}
\end{prop}
\begin{proof}
Start from the definition \eqref{eq:def_f_divergence}
\begin{equation}
\begin{aligned}
    D_f(\rho \Vert \sigma) &= \int_1^\infty f^{\prime \prime}(\gamma) \operatorname{tr}^+[\rho - \gamma\sigma] \\
    &\hspace{1cm} + \gamma^{-3}f^{\prime \prime}(\gamma ^{-1})\operatorname{tr}^+[\sigma - \gamma \rho] \ d\gamma,
\end{aligned}
\end{equation}
With the change of variables $s=\gamma^{-1}$ in the second term,
\begin{equation}
\begin{aligned}
D_f(\rho\Vert\sigma)
&=\int_{1}^{\infty} f^{\prime \prime}(s) \tr^+[\rho - s \sigma] \ ds \\
&\hspace{1cm}+\int_{0}^{1} f^{\prime \prime}(s) \tr^-[\rho-s\sigma] \ ds.
\end{aligned}
\end{equation}
If $\mu\sigma\leq\rho\leq\lambda\sigma$, then $\tr^+[\rho-s \sigma]=0$ for $s\geq \lambda$ and 
$\tr^-[\rho-s\sigma]=0$ for $s\leq \mu$, hence
\begin{align}
D_f
=\int_{\mu}^{1} f^{\prime \prime}(s) \tr^-[\rho-s\sigma] \ ds
+\int_{1}^{\lambda} f^{\prime \prime}(s) \tr^+[\rho - s\sigma] \ ds.
\end{align}
Using $\tr^+[\rho-s\sigma]
=\tr^-[\rho-s\sigma]+(1-s)$ gives
\begin{align}
D_f
=\int_{\mu}^{\lambda} f^{\prime \prime}(s) \tr^-[\rho-s\sigma] \ ds
+\int_{1}^{\lambda} (1-s)f^{\prime \prime}(s) \ ds.
\end{align}
Integration by parts yields
\begin{equation}
\begin{aligned}
\int_{1}^{\lambda} (1-s)f^{\prime \prime}(s) ds
&=\big[f(s)+(1-s)f^{\prime }(s)\big]_{1}^{\lambda} \\
&=f(\lambda)+(1-\lambda)f^{\prime }(\lambda),
\end{aligned}
\end{equation}
which proves \eqref{eq:fd_negpart}. 
\end{proof}

\autoref{prop:fd_general_formula} shows that all tools needed for the relative entropy program \eqref{eq:general_problem} and the special case of $f(s) = x\log x$ can be discussed in the more general class of $f$-divergences and the optimization problem 
\begin{align}\label{eq:general_problem_f}
    c_f \coloneqq  \inf & \ D_f(\rho \Vert \sigma)  \\ \nonumber
    \operatorname{s.t.} \ & h_i(\rho,\sigma) \geq 0 \quad i = 1, \dots, n \\ \nonumber
     &\mu \sigma \leq \rho \leq \lambda \sigma \\ \nonumber
    &\sigma, \rho \in \mathcal{S}(\mathcal{H}) .
\end{align}
Particularly, the estimates in \eqref{eq:relaxation_lower_bound} and \eqref{eq:relaxation_upper_bound} are straightforward to generalize.   

We divide the convergence analysis of \eqref{eq:relaxation_lower_bound} and \eqref{eq:relaxation_upper_bound} into two parts, namely the analysis of $\tr^+[\sigma s - \rho]$, which becomes a central ingredient and the convergence analysis itself then becomes the second part of this section. 

\textit{-- Analysis of $\tr^+[\sigma s - \rho]$ :}  

Our method for the relaxations in \eqref{eq:relaxation_lower_bound} and \eqref{eq:relaxation_upper_bound} is based on the following observation in the $f$-divergence setting. 
Denote $w(s)\coloneqq f''(s)\geq   0$ (recall that for a convex, differentiable function the derivative is an increasing function \cite[Thm. 1.4.3]{Niculescu2025} and thus the second derivative is positive) and use the terminology from \autoref{prop:fd_general_formula} with $\operatorname{const}(\lambda) \coloneqq f(\lambda)+(1-\lambda)f'(\lambda)$, then we have
\begin{equation}
\begin{aligned}
    D_f(\rho\Vert\sigma) - \operatorname{const}(\lambda)
    &= \int_{\mu}^{\lambda} w(s) \tr^+[\sigma s-\rho] \ ds \label{eq:jencova_integral_f}\\
    &= \int_{\mu}^{\lambda} w(s) \sup_{0\leq P\leq\mathds{1}}\tr\!\big[P(\sigma s-\rho)\big] \ ds. 
\end{aligned}
\end{equation}
At this point, one may evaluate the supremum in $P$ pointwise in $s\in[\mu,\lambda]$. 
But since $w(s)\geq   0$, it is valid for obtaining lower bounds to estimate the supremum once \emph{after} integration. For this purpose choose $a,b \in [\mu,\lambda]$ such that $\mu \leq a< b \leq \lambda$ and estimate
\begin{equation}\label{eq:expl_lower_bounds}
\begin{aligned}
   & \int_{a}^{b} w(s) \sup_{0\leq P\leq\mathds{1}}\tr\!\big[P(\sigma s-\rho)\big] \ ds \\
    &\hspace{2cm} \ge\ \sup_{0\leq P\leq \mathds{1}} \int_a^b w(s) \tr\!\big[P(\sigma s-\rho)\big] \ ds.
\end{aligned}
\end{equation}
Of course, this is in general a loose estimate, but its interpretation is that we choose the best \emph{linear} functional (via a single effect $P$) that lower bounds the trace term \emph{at the level of the weighted integral}, rather than approximating the pointwise convex function $s\mapsto \tr^+[\sigma s-\rho]$ itself. 
Thus the supremum is reinterpreted from a pointwise optimization to an optimization over integrated values, while exploiting the structural properties of $s\mapsto \tr^+[\sigma s-\rho]$ collected in the following \autoref{lem:properties_g}. 

\begin{lem}[Properties of Divergence]\label{lem:properties_g}
    Let $\rho,\sigma \in \mathcal{S}(\mathcal{H})$ be two quantum states. Then $g(s) \coloneqq \operatorname{tr}^+[\sigma s - \rho]$ has the following properties
    \begin{enumerate}
        \item[(a)] $g$ is convex for $s \in \mathbb{R}$ and in particular continuous in $(s,\rho,\sigma)$.
        \item[(b)] $g$ is monotonically increasing. 
        \item[(c)] $\tr^+[\sigma s - \rho]$ satisfies the data processing inequality, i.e. for every positive, trace-nonincreasing channel $\Phi$ we have
        \begin{align}
            \tr^+[\sigma s - \rho] \geq \tr^+[\Phi(\sigma) s - \Phi(\rho)] \quad \rho,\sigma \in \mathcal{S}(\mathcal{H}).
        \end{align}
        \item[(d)] for all $s\in \mathbb{R}$ we have $s-1 \leq \operatorname{tr}^+[\sigma s - \rho] \leq s$.
    \end{enumerate}
\end{lem}
\begin{proof}
    Supplementary material: \autoref{appendix_proof_lemma}.
\end{proof}

The idea of our technique is to use a grid $\mathbf{t}$ and apply \eqref{eq:expl_lower_bounds} interval-wise.
Combining the facts that $\tr^+[\sigma s - \rho]$ is convex and monotonically increasing, and that interchanging the integration and supremum yields valuable lower bounds, suggests—at least heuristically—that even a small number of grid points is sufficient to obtain non-trivial lower bounds on \eqref{eq:general_problem_f}. 

\textit{-- Detailed Convergence Analysis :}  

From~\autoref{fig:convergence} it is geometrically evident that the optimised lower bound in~\autoref{prop:mainprop} has an error no greater than that of the upper bound: by convexity, the mirrored straight line $g^{\prime\prime}$ is a feasible supporting line (see the discussion below for a precise definition of supporting line) for the lower bound in~\eqref{eq:relaxation_lower_bound}, and thus we obtain at least the same error dependence as the convex interpolation in the upper bound. Without loss of generality, and in order to focus the convergence analysis on the upper bound, in the following we formalize this geometric observation using the properties of convex functions.

For $s\in\mathbb{R}$ let
\begin{align}
g(s)\coloneqq \tr^+[\sigma s-\rho],\qquad
w(s)\coloneqq f''(s)\ge 0,\quad f(1)=0,
\end{align}
and fix an interval $[a,b]\subseteq[\mu,\lambda]$. By \autoref{lem:properties_g}(a), $g$ is convex.

Recall the \emph{subdifferential} of $g$ at $s_0$:
\begin{align}
\partial g(s_0)\coloneqq \{\,m\in\mathbb{R}: \ g(s)\ge g(s_0)+m(s-s_0)\ \text{for all } s\in\mathbb{R}\,\}.
\end{align}
For any $m\in\partial g(s_0)$ the affine map
\begin{align}\label{eq:supporting_lines}
T_{s_0,m}(s)\coloneqq g(s_0)+m(s-s_0)
\end{align}
is a supporting line of $g$ at $s_0$ (so $T_{s_0,m}\le g$ on $\mathbb{R}$ and $T_{s_0,m}(s_0)=g(s_0)$).
The subdifferential mapping is \emph{monotone} (see e.g. \cite[Thm. 4.3.12]{Niculescu2025}): if $s_1<s_2$, $m_1\in\partial g(s_1)$ and $m_2\in\partial g(s_2)$, then
\begin{equation}\label{eq:monotonicity_subgradients}
0\le (m_2-m_1)(s_2-s_1)\quad\Rightarrow\quad m_1\le m_2.
\end{equation}

Define the secant of $g$ over $[a,b]$ by\footnote{Recall the notion of left and right derivatives, denoted as $g'_+(\cdot),g'_-(\cdot)$, for convex functions from e.g. \cite[Thm. 1.4.2]{Niculescu2025}}
\begin{equation}\label{eq:def_secant}
\begin{aligned}
S_{a,b}(s)&\coloneqq g(a)+m_{a,b}(s-a),\\
m_{a,b}&\coloneqq \frac{g(b)-g(a)}{b-a}\in\bigl[g'_+(a),\, g'_-(b)\bigr].
\end{aligned}
\end{equation}
By convexity and the generalized mean–value theorem for convex functions (see e.g. \cite[Chap. 2, Ex. 2]{Niculescu2025}), there exists $s^\star\in[a,b]$ with
$m_{a,b}\in\partial g(s^\star)$. Hence the particular supporting line we use is simply
\begin{align}
T_{s^\star,m_{a,b}}(s)=g(s^\star)+m_{a,b}(s-s^\star)\le g(s)\quad\text{for all }s,
\end{align}
and $T_{s^\star,m_{a,b}}(s^\star)=g(s^\star)$.

Moreover, since $g$ is the pointwise supremum of the affine forms $s\mapsto \tr[P(\sigma s-\rho)]$ over effects $P$ ($0\le P\le I$), the supremum at $s^\star$ is attained by at least one effect $P^\star$. For such a $P^\star$ we have
\begin{align}
g(s^\star)=\tr\!\bigl[P^\star(\sigma s^\star-\rho)\bigr]\quad\text{and}\quad
m_{a,b}=\tr[P^\star\sigma],
\end{align}
so, equivalently, $T_{s^\star,m_{a,b}}(s)=\tr\!\bigl[P^\star(\sigma s-\rho)\bigr]$.

As a minor result, we require an adapted error analysis for the trapezoid method for the special case of a convex function. This result is essentially the classical error estimate for the trapezoid rule, as found in standard textbooks on numerical analysis (see, e.g., \cite[Eq.~5.1.7]{Stewart2022}), but stated without the assumption of differentiability.
 
\begin{lem}\label{lem:convergence_trapezoid}
  Let $g$ be convex on $[\mu,\lambda]$ and let 
  $\mu = t_1 < t_2 < \cdots < t_r = \lambda$ be a partition.  Denote on each $[t_k,t_{k+1}]$ the secant from \eqref{eq:def_secant} as
  \begin{align}
    S_k(t) \equiv S_{t_k,t_{k+1}}(t).
  \end{align}
  Then the trapezoidal‐rule error satisfies
  \begin{align}\label{eq:lem_convexity_trapezoid}
    &\sum_{k=1}^{r-1}\int_{t_k}^{t_{k+1}}S_k(t) \ dt -\int_\mu^\lambda g(t) \ dt
    \leq\\
    &\hspace{2cm} \sum_{k=1}^{r-1}\frac{(t_{k+1}-t_k)^2}{8}\bigl(g^\prime_-(t_{k+1})-g^\prime_+(t_k)\bigr).
  \end{align}
\end{lem}
\begin{proof}
    See \autoref{proof_convergence}.
\end{proof}

\autoref{lem:convergence_trapezoid} enables us to prove the following technical lemma.
\begin{lem}\label{lem:lower_bounds_at_least_upper_bounds}
For every convex function $g$ and any $a<b$, the secant $S_{a,b}$ of $g$ on $[a,b]$ and any supporting line $T_{a,b}$ of $g$ at some $s^\star\in[a,b]$ and $w\ge 0$ and $L_{a,b}\coloneqq \sup_{s\in[a,b]} w(s)$, we have
\begin{align}
\int_a^b \! w(s)\,\bigl(S_{a,b}(s)-g(s)\bigr)\,ds   \le   
\frac{L_{a,b}}{8}\,(b-a)^2\,\bigl(g'_-(b)-g'_+(a)\bigr),\label{eq:wupper_local}\\
\int_a^b \! w(s)\,\bigl(g(s)-T_{a,b}(s)\bigr)\,ds   \le   
\frac{L_{a,b}}{8}\,(b-a)^2\,\bigl(g'_-(b)-g'_+(a)\bigr).\label{eq:wlower_local}
\end{align}
\end{lem}
\begin{proof}
\autoref{lem:lower_bounds_at_least_upper_bounds}    
\end{proof}

Thus, on each grid interval $[t_k,t_{k+1}]$:
\begin{itemize}
\item[(i)] the \emph{upper} error from convex interpolation by secants, and
\item[(ii)] the \emph{lower} error from the best affine minorant realized by a single effect $P$
\end{itemize}
are both bounded by the common quantity in \eqref{eq:wupper_local}–\eqref{eq:wlower_local}.
Thus, using the coarse weights $L_k\coloneqq\sup_{s\in[t_k,t_{k+1}]} w(s)$, the optimized lower bound is \emph{at least as good as} the upper bound.

Moreover, equation \eqref{eq:lem_convexity_trapezoid} has the interesting consequence that for convex $g$, $g'_+$ and $g'_-$ exist everywhere and are nondecreasing, hence
\begin{align}\label{eq:upper_bound_on_derivatives}
\sum_{k=1}^{r-1}\!\bigl(g'_-(t_{k+1})-g'_+(t_k)\bigr)
\le g'_-(t_r)-g'_+(t_1),
\end{align}
by the observation that $-(g_+'(t_k)-g_-'(t_k))\leq 0$.

\begin{prop}[Convergence for $f$-divergences]\label{prop:convergence_f_div}
Let $\rho,\sigma\in\mathcal{S}(\mathcal{H})$ with $\mu\sigma\leq\rho\leq\lambda\sigma$. 
Approximate a $f$-divergence via the convex interpolation upper bound on a grid 
$\mathbf{t}=(\mu=t_1<t_2<\cdots<t_r=\lambda)$:
\begin{align}
\sum_{k=1}^{r-1}\int_{t_k}^{t_{k+1}} w(s) g(s) ds
\ \le\ \sum_{k=1}^{r-1}\int_{t_k}^{t_{k+1}} w(s) S_k(s) ds,
\end{align}
with $g(s)=\tr^+[\sigma s-\rho]$ and $S_k$ the secant of $g$ on $[t_k,t_{k+1}]$. 
Fix a target accuracy $\varepsilon>0$ and choose the grid recursively by
\begin{equation}\label{eq:def_special_grid_f}
t_k = 
\begin{cases}
    \mu & k=1 \\
    t_{k-1} + \sqrt{\frac{8 \varepsilon}{L_{k-1}}} & k\geq 2 
\end{cases}
\end{equation}
with $L_{k-1}\coloneqq \sup_{s\in[t_{k-1}, t_k]} w(s)$. Then the total interpolation error obeys
\begin{align}
\sum_{k=1}^{r-1}\int_{t_k}^{t_{k+1}} w(s) \bigl(S_k(s)-g(s)\bigr) ds\ \le\ \varepsilon,
\end{align} 
and the number of grid points satisfies
\begin{align}
 r   =   \mathcal{O}\!\Big(\tfrac{1}{\sqrt{\varepsilon}}\int_{\mu}^{\lambda}\!\sqrt{f''(s)} ds\Big).
\end{align}
\end{prop}

\begin{proof}
On each interval we have (see \autoref{lem:convergence_trapezoid})
\begin{equation}
\begin{aligned}
&\int_{t_k}^{t_{k+1}} w(s) \bigl(S_k(s)-g(s)\bigr) ds \\
&\hspace{1cm} \le\ \frac{L_k}{8} (t_{k+1}-t_k)^2 \bigl(g'_-(t_{k+1})-g'_+(t_k)\bigr).
\end{aligned}
\end{equation}
Choose the grid by \eqref{eq:def_special_grid_f}, so that $(t_{k+1}-t_k)^2\leq8\varepsilon/L_k$; hence the $k$-th interval contributes at most 
$\varepsilon \bigl(g'_-(t_{k+1})-g'_+(t_k)\bigr)$. 
Summing over $k$ and using \eqref{eq:upper_bound_on_derivatives} with \autoref{lem:properties_g} (d), which yields that \eqref{eq:upper_bound_on_derivatives} can be estimated with $1$ for $g(s) = \tr^+[\sigma s -\rho]$, because $\operatorname{tr}[\sigma]= 1$, implies
\begin{equation}
\begin{aligned}
\sum_{k=1}^{r-1}\int_{t_k}^{t_{k+1}} w(s) \bigl(S_k-g\bigr) ds
 &\leq\varepsilon\sum_{k=1}^{r-1}\bigl(g'_-(t_{k+1})-g'_+(t_k)\bigr) \\
 &\leq \varepsilon.
\end{aligned}
\end{equation}

For the grid size, from \eqref{eq:def_special_grid_f} we have
$t_{k}-t_{k-1}=\sqrt{8\varepsilon/L_{k-1}}$. By continuity of $w$, there exists $s_{k-1}\in[t_{k-1},t_k]$ with $L_{k-1}=w(s_{k-1})$, hence
\begin{align}
1\ \le\ \frac{\sqrt{w(s_{k-1})}}{\sqrt{8\varepsilon}}(t_k-t_{k-1}).
\end{align}
Summing from $k=2$ to $r$ and passing to the Riemann sum limit gives
\begin{equation}
\begin{aligned}
r-1  &\leq\frac{1}{\sqrt{8 \varepsilon}} \sum_{k=1}^{r-1}\sqrt{w(s_k)} (t_{k+1}-t_k) \\
 &\leq\frac{1}{\sqrt{8 \varepsilon}}\int_{\mu}^{\lambda}\sqrt{w(s)} ds,
\end{aligned}
\end{equation}
which proves the stated bound on $r$. 
\end{proof}

Recalling now the general optimization problem in the language of $f$-divergences \eqref{eq:general_problem_f} yields that we can approximate the value $c_f$ by the upper bounds $c_u(\mathbf{t})$, because the bounds guarantee uniform convergence as stated in the next corollary.  
\begin{corr}\label{cor:value_convergence}
Let $c_f$ be the value of \eqref{eq:general_problem}.
For a grid $\mathbf t=(\mu=t_1<\cdots<t_r=\lambda)$ define
\begin{equation}
\begin{aligned}
&U_{\mathbf t}(\rho,\sigma)\;:=\;\sum_{k=1}^{r-1}\int_{t_k}^{t_{k+1}} w(s)\,S_k(s)\,ds \\
&\hspace{3cm}\;+\;f(\lambda)+(1-\lambda)f'(\lambda),
\end{aligned}
\end{equation}
and the corresponding relaxation value
\begin{equation}
\begin{aligned}
&c_u(\mathbf t) = \\
&\hspace{0.5cm}\inf\Big\{\,U_{\mathbf t}(\rho,\sigma):
\ h_i(\rho,\sigma)\ge 0,\ \mu\sigma\le\rho\le\lambda\sigma,\ \rho,\sigma\in\mathcal S(\mathcal H)\Big\}.
\end{aligned}
\end{equation}
Similarly, define with \eqref{eq:supporting_lines} 
\begin{equation}
\begin{aligned}
&V_{\mathbf t}(\rho,\sigma)\;:=\;\sum_{k=1}^{r-1}\int_{t_k}^{t_{k+1}} w(s)\,T_{t_k,t_{k+1}}(s)\,ds \\
&\hspace{3cm} \;+\;f(\lambda)+(1-\lambda)f'(\lambda),
\end{aligned}
\end{equation}
and the corresponding relaxation value
\begin{equation}
\begin{aligned}
&c_l(\mathbf t) = \\
&\hspace{0.5cm}\inf\Big\{\,V_{\mathbf t}(\rho,\sigma):
\ h_i(\rho,\sigma)\ge 0,\ \mu\sigma\le\rho\le\lambda\sigma,\ \rho,\sigma\in\mathcal S(\mathcal H)\Big\}.
\end{aligned}
\end{equation}
If $\mathbf t$ is chosen by \eqref{eq:def_special_grid_f} for a target accuracy $\varepsilon>0$, then
\begin{align}
c_l(\mathbf{t}) - \varepsilon\leq c_f\ \le\ c_u(\mathbf t)\ \le\ c_f+\varepsilon
\end{align}
and 
\begin{align}
    r=O\!\Big(\tfrac{1}{\sqrt{\varepsilon}}\int_{\mu}^{\lambda}\!\sqrt{f''(s)}\,ds\Big).
\end{align}
In particular, for any refining sequence of grids with $\varepsilon\downarrow 0$ we have  $c_l(\mathbf t)\uparrow c_f$ and $c_u(\mathbf t)\downarrow c_f$ (monotone value convergence).
\end{corr}

\begin{proof}
By Proposition~\ref{prop:convergence_f_div} and \autoref{lem:lower_bounds_at_least_upper_bounds}, for every feasible pair $(\rho,\sigma)$,
\begin{equation}
\begin{aligned}
&D_f(\rho\Vert \sigma)  - \varepsilon \leq V_{\mathbf{t}}(\rho,\sigma)\leq D_f(\rho\Vert\sigma)  \\
&\hspace{3cm}\leq U_{\mathbf t}(\rho,\sigma)\ \le\ D_f(\rho\Vert\sigma)+\varepsilon.
\end{aligned}
\end{equation}
Taking the infimum over the common feasible set yields (similarly for the lower bound)
\begin{align}
c_f=\inf D_f\ \le\ \inf U_{\mathbf t}=c_u(\mathbf t)\ \le\ \inf(D_f+\varepsilon)=c_f+\varepsilon.
\end{align}
The stated bound on $r$ follows directly from Proposition~\ref{prop:convergence_f_div}. 
\end{proof}

\section{Acknowledgements}
We thank Mario Berta, Tobias J. Osborne, Hermann Kampermann, Martin Plavala and Zhen-Peng Xu for fruitful discussions. 
We thank Omar Fawzi for pointing us to the reference \cite{Fawzi_2018} from which we took the examples in the supplementary material \autoref{sec:quantum_shannon}. We thank Florian Oerke and Felix Golke for spotting several typos in the manuscript. 
GK acknowledges support from the Excellence Cluster - Matter and Light for Quantum Computing (ML4Q). GK acknowledges funding by the European Research Council (ERC Grant Agreement No. 948139).
R.S.\ is supported  by the DFG under Germany's Excellence Strategy - EXC-2123 QuantumFrontiers - 390837967 and  SFB 1227 (DQ-mat), the Quantum Valley Lower Saxony, and the BMBF projects ATIQ, SEQUIN, QuBRA and CBQD.  

\section{Code availability}
 An implementation for all use cases is accessible in the \href{https://github.com/gereonkn/relative-entropy-optimization.git}{repository}.  
\bibliography{main}
\section*{Author contribution}
GK and RS contributed equally to the work. 
\section*{Competing Interests}
All authors declare no financial or non-financial competing interests.

\begin{widetext}

\appendix

\section{Proof of \texorpdfstring{\autoref{prop:mainprop}}{}}\label{appendix_proof_propmain}
The first statement follows from the following calculation. 
Let $\mu = t_1 \leq \cdots \leq t_r = \lambda$. Then we have 
\begin{align*}
    \int_\mu^\lambda \frac{ds}{s} \, \operatorname{tr}^+[\sigma s - \rho] 
    &= \sum_{k=1}^{r-1} \int_{t_k}^{t_{k+1}} \frac{ds}{s} 
       \sup_{0 \leq P \leq \mathbb{1}} \operatorname{tr}\big[P(\sigma s - \rho)\big] \\
    &\geq \sum_{k=1}^{r-1} \sup_{0 \leq P \leq \mathbb{1}} 
       \int_{t_k}^{t_{k+1}} \frac{ds}{s} \, \operatorname{tr}\big[P(\sigma s - \rho)\big] \\
    &= \sum_{k=1}^{r-1} \sup_{0 \leq P \leq \mathbb{1}} 
       \Big[ \operatorname{tr}(P\sigma) \underbrace{(t_{k+1}-t_k)}_{=:\beta_k}
       + \operatorname{tr}(P\rho) \underbrace{\log \frac{t_k}{t_{k+1}}}_{=:\alpha_k} \Big] \\
    &= \sum_{k=1}^{r-1} \sup_{0 \leq P \leq \mathbb{1}} 
       \operatorname{tr}\big[ P(\sigma \beta_k + \rho \alpha_k) \big] \\
    &= \inf_{\mu_1, \ldots, \mu_{r-1}} 
       \sum_{k=1}^{r-1} \operatorname{tr}[\mu_k] \\
    &\quad\quad \operatorname{s.t.} \quad \mu_k \geq \alpha_k \rho + \beta_k \sigma.
\end{align*}
This gives us~\eqref{eq:relaxation_lower_bound}.
\section{Lemma for lower bounds}\label{lem:Lower_bounds}
\begin{lem}\label{lem:rewriting_with_complementary_device_dependent}
    Assume a projective measurement given by a channel 
    \begin{align}
        \Phi_{i^\star}^A:\mathcal{S}(\mathcal{H}_A) \to \mathcal{S}(\mathcal{X}_{i^\star}).
    \end{align}
    For a state $\rho_{AB}\in \mathcal{S}(\mathcal{H}_A \otimes \mathcal{H}_B)$ with purification $\rho_{AB}\in \mathcal{S}(\mathcal{H}_A \otimes \mathcal{H}_B \otimes \mathcal{H}_E)$ we have 
    \begin{align}
        H(\mathcal{X}_{i^\star} \vert E)_{\Phi_{i^\star}^A \otimes \operatorname{id}_E(\rho_{AE})} = D\big(\rho_{AB} \Vert \big(\Phi_{i^\star}^A\big)^\prime \otimes \operatorname{id}_B(\rho_{AB})\big),
    \end{align}
    whereby $(\Phi_{i^\star}^A\big)^\prime $ denotes the complementary channel of $\Phi_{i^\star}^A$.
\end{lem}
\begin{proof}
    We consider the Stinespring-dilation of the channel $\Phi_{i^\star}^A $ given by a unitary
    \begin{align}
        U:\mathcal{H}_A   \otimes \mathcal{H}_M \to \mathcal{H}_A   \otimes \mathcal{H}_M.  
    \end{align}
    Now, by the definition of the conditional von Neumann entropy, 
    \begin{equation}
    \begin{aligned}
        H(\mathcal{X}_{i^\star} \vert E)_{\Phi_{i^\star}^A \otimes \operatorname{id}_E(\rho_{AE})}  &= H(\mathcal{X}_{i^\star}  E)_{\Phi_{i^\star}^A \otimes \operatorname{id}_E(\rho_{AE})} - H(E)_{\rho_E} \\
        &= H(MB)_{\operatorname{tr}_{AE}[U\rho_{ABEM}U^\dagger]}
    \end{aligned}
    \end{equation}
    and by the fact that the entropy equally marginalizes for pure states. Similarly, we can do this for $H(E)_{\rho_E}$ such that we get
    \begin{equation}
        \begin{aligned}
            \ldots &= H(MB)_{\operatorname{tr}_{AE}[U\rho_{ABEM}U^\dagger]} - H(AB)_{\rho_{AB}} \\
            & = H(MB)_{\big(\Phi_{i^\star}^A\big)^\prime(\rho_{AB})} - H(AB)_{\rho_{AB}}
        \end{aligned}
    \end{equation}
    with the definition of the complementary channel. Furthermore, we have
    \begin{equation}\label{eq:proof_identification_proof_complementary_channel}
    \begin{aligned}
        H(MB)_{\big(\Phi_{i^\star}^A)^\prime (\rho_{AB})} &= -\operatorname{tr}[\big(\Phi_{i^\star}^A)^\prime (\rho_{AB}) \log \big(\Phi_{i^\star}^A)^\prime (\rho_{AB})] \\
        &=-\operatorname{tr}[\rho_{AB} \bigg( \big(\Phi_{i^\star}^A)^\prime\bigg)^\star\bigg( \log \big(\Phi_{i^\star}^A)^\prime (\rho_{AB})\bigg)] \\
        &= -\operatorname{tr}[\rho_{AB}  \log \big(\Phi_{i^\star}^A)^\prime (\rho_{AB})]
    \end{aligned}
    \end{equation}
    In the last step we added that the measurement channel $\Phi_{i^\star}^A$ is projective, we can identity it with its complementary channel and similarly with its dual. Furthermore, projective channels have the property that $\big(\Phi_{i^\star}^A\big)^2 = \Phi_{i^\star}^A$. With \eqref{eq:proof_identification_proof_complementary_channel} we have
    \begin{equation}
        \begin{aligned}
            \ldots &= H(MB)_{ \big(\Phi_{i^\star}^A)^\prime(\rho_{AB})} - H(AB)_{\rho_{AB}} \\
            &= D(\rho_{AB} \Vert \big(\Phi_{i^\star}^A)^\prime(\rho_{AB})).
        \end{aligned}
    \end{equation}
\end{proof}

\section{Derivation Upper Bound}\label{appendix_upper_bound}
Choose a grid $\mu = t_1 \leq \cdots \leq t_r = \lambda$ and define
\begin{align*}
    y_k \coloneqq \sup_{0\leq P \leq \mathds{1}} \tr[P (\sigma t_k - \rho)] 
\end{align*}
Then we have due to the convexity of $\tr^+[\sigma s - \rho]$
\begin{align*}
    \int_\mu^\lambda \frac{ds}{s}\tr^+[\sigma s - \rho] &= \sum_{k=1}^{r-1} \int_{t_k}^{t_{k+1}} \frac{ds}{s}\tr^+[\sigma s - \rho] \\
    &\leq \sum_{k=1}^{r-1} \int_{t_k}^{t_{k+1}} \frac{ds}{s} (\frac{y_{k+1} - y_k}{t_{k+1}-t_k}(s-t_k) + y_k)  \\
    &= \sum_{k=1}^{r-1} (y_{k+1}- y_k) \ + (y_k - t_k \frac{y_{k+1} - y_k}{t_{k+1}-t_k})\log \frac{t_{k+1}}{t_k} \\
    &= \sum_{k=1}^{r-1} y_{k+1} - t_k \frac{y_{k+1}}{t_{k+1}-t_k} \log \frac{t_{k+1}}{t_k}  + \sum_{k=1}^{r-1} (y_k + \frac{t_k y_k}{t_{k+1}-t_k}) \log \frac{t_{k+1}}{t_k} - y_k \\
    &= (y_1 + \frac{t_1 y_1}{t_2 - t_1})\log \frac{t_2}{t_1} - y_1  + y_r - \frac{t_{r-1} y_r}{t_r - t_{r-1}}\log \frac{t_r}{t_{r-1}} + \sum_{k=1}^{r-2} y_{k+1} - t_k \frac{y_{k+1}}{t_{k+1}-t_k} \log \frac{t_{k+1}}{t_k} \\
    &\quad + \sum_{k=2}^{r-1} (y_k + \frac{t_k y_k}{t_{k+1}-t_k}) \log \frac{t_{k+1}}{t_k} - y_k \\
    &= (y_1 + \frac{t_1 y_1}{t_2 - t_1})\log \frac{t_2}{t_1} - y_1  + y_r - \frac{t_{r-1} y_r}{t_r - t_{r-1}}\log \frac{t_r}{t_{r-1}} + \sum_{k=2}^{r-1} \cancel{y_{k}} - t_{k-1} \frac{y_{k}}{t_{k}-t_{k-1}} \log \frac{t_{k}}{t_{k-1}} \\
    &\quad + \sum_{k=2}^{r-1} (y_k + \frac{t_k y_k}{t_{k+1}-t_k}) \log \frac{t_{k+1}}{t_k} \cancel{- y_k} \\ 
    &= y_1[(1 + \frac{t_1}{t_2 - t_1})\log \frac{t_2}{t_1} - 1]  + y_r[ 1 - \frac{t_{r-1}}{t_r - t_{r-1}}\log \frac{t_r}{t_{r-1}}] \\ 
    &\quad + \sum_{k=2}^{r-1} y_k[(1+ \frac{t_k}{t_{k+1}-t_k})\log  \frac{t_{k+1}}{t_k}  - \frac{t_{k-1}}{t_k - t_{k-1}}\log \frac{t_k}{t_{k-1}}].
\end{align*}

Inserting the definition of $y_k$ yields the following definitions for $\delta_k$ and $\gamma_k$
\begin{align*}
    \delta_k = \begin{cases}
        [(1 + \frac{t_1}{t_2 - t_1})\log \frac{t_2}{t_1} - 1]\cdot t_1 & k = 1 \\
        [ 1 - \frac{t_{r-1}}{t_r - t_{r-1}}\log \frac{t_r}{t_{r-1}}]\cdot t_r & k = r \\
        [(1+ \frac{t_k}{t_{k+1}-t_k})\log  \frac{t_{k+1}}{t_k}  - \frac{t_{k-1}}{t_k - t_{k-1}}\log \frac{t_k}{t_{k-1}}]\cdot t_k & \text{else}.
    \end{cases}
\end{align*}
and for $\delta_k$
\begin{align*}
    \gamma_k = \begin{cases}
        -[(1 + \frac{t_1}{t_2 - t_1})\log \frac{t_2}{t_1} - 1] & k = 1 \\
        -[ 1 - \frac{t_{r-1}}{t_r - t_{r-1}}\log \frac{t_r}{t_{r-1}}] & k = r \\
        -[(1+ \frac{t_k}{t_{k+1}-t_k})\log  \frac{t_{k+1}}{t_k}  - \frac{t_{k-1}}{t_k - t_{k-1}}\log \frac{t_k}{t_{k-1}}] & \text{else}.
    \end{cases}
\end{align*}
Thus we conclude
\begin{align}\label{eq:coeff_upper}
    \int_\mu^\lambda \frac{ds}{s}\tr^+[\sigma s - \rho] & \leq \sum_{k=1}^r \sup_{0\leq P \leq \mathds{1}}\tr[P(\gamma_k \rho + \delta_k \sigma)] \\
    &= \inf_{\nu_1,\ldots,\nu_r}\sum_{k=1}^r \tr[\nu_k] \\
    &\quad \quad \operatorname{s.t.} \ \nu_k \geq \gamma_k \rho + \delta_k \sigma \\
    &\hspace{1.5cm} \nu_k \geq 0.
\end{align}

\section{Proof of \texorpdfstring{\autoref{lem:properties_g}}{}}\label{appendix_proof_lemma}
\begin{proof}
    \textit{-- Convexity and Continuity:}

The convexity follows by the following calculation similar to \cite{Hirche2023}
\begin{align*}
    \tr^+[\sigma (\lambda s_1 + (1-\lambda)s_2) - \rho] &= \tr^+[\lambda (\sigma s_1 - \rho) + (1-\lambda) (\sigma s_2 - \rho) ] \\
    &= \sup_{0\leq P \leq \mathds{1}} \lambda \tr[P (\sigma s_1 - \rho)] + (1-\lambda)\tr[P (\sigma s_2 - \rho) ] \\
    &\leq \lambda \sup_{0\leq P \leq \mathds{1}} \tr[P (\sigma s_1 - \rho)] + (1-\lambda)\sup_{0\leq P \leq \mathds{1}}\tr[P (\sigma s_2 - \rho) ] \\
    &= \lambda \tr^+[\sigma s_1 -\rho] + (1-\lambda) \tr^+[\sigma s_2 - \rho].
\end{align*}
For continuity let $\rho,\tilde{\rho},\sigma,\tilde{\sigma}$ be quantum states and $s_1,s_2 \in \mathbb{R}_{\geq 0}$. Then we have similar to \cite{Sharma_2013}
\begin{align*}
    2\cdot \vert \tr^+[\sigma s_1 - \rho] - \tr^+[\tilde{\sigma} s_2 - \tilde{\rho} ]\vert &= \vert \Vert s_1 \sigma - \rho \Vert_1 + (s_1-1) - (\Vert \tilde{\sigma} s_2  - \tilde{\rho} \Vert_1 + s_2 - 1)\vert \\
    &\leq \vert s_2 - s_1 \vert + \Vert \rho - \tilde{\rho}\Vert_1 + \Vert s_1 \sigma - s_2 \tilde{\sigma}\Vert_1.
\end{align*}
    whereby we used the reversed triangle inequality for the one-norm. In particular we see that the divergence is continuous in the states. In the other way around for equal states and different $s_{1/2}$ we have continuity in $s$. \\
\textit{-- Monotony:}
The monotony is a direct consequence of the fact that for $s_1\geq s_2$ we have
\begin{align*}
    \sigma s_1 - \rho \geq \sigma s_2 - \rho.
\end{align*}
in operator inequality. 

\textit{-- data processing inequality:} Consider a channel $\Phi:\mathcal{S}(\mathcal{H}) \to \mathcal{S}(\mathcal{H})$. We assume that $D_{\text{max}}(\rho \Vert \sigma) < \infty$ and satisfies the data processing inequality. This yields that
$D(\rho \Vert  \sigma) - D(\Phi(\rho) \Vert \Phi(\sigma)) $ is a finite number. Adapting from \cite[Lem. 4]{Sharma_2013} the proof that $\tr^+[\sigma s - \rho]$ satisfies the data processing inequality, namely let $\sigma s - \rho = Q - S$ with $Q,S\geq 0$. Then let $P\coloneqq P_{\Phi(\sigma)s - \Phi(\rho)\geq 0}$ and
\begin{align*}
    \tr^+[\sigma s - \rho] &= \tr[Q] \\
                        &= \tr[\Phi(Q)] \\
                        &\geq \tr[P(\Phi(Q) - \Phi(S))] \\
                        &= \tr[P(\Phi(\sigma s - \rho)] \\
                        &= \tr^+[\Phi(\sigma) s - \Phi(\rho)].
\end{align*}        
Therefore we know that for any $\kappa \geq \lambda$
\begin{align*}
    0 &= \int_{\lambda}^{\kappa} \frac{ds}{s} \underbrace{\tr^+[\sigma s - \rho] - \tr^+[\Phi(\sigma)s  - \Phi(\rho)]}_{\geq 0} \\
    &= \int_{\lambda}^{\kappa} \frac{ds}{s} \vert \tr^+[\sigma s - \rho] - \tr^+[\Phi(\sigma)s  - \Phi(\rho)] \vert. 
\end{align*}
The property now follows from the fact that $\Vert \cdot \Vert_1$ is a norm on the space of integrable functions and we know easily that every classical dichotomy of states satisfies the property asymptotically and the fact that channels are trace preserving.

\textit{-- Upper and lower bounds:}

For the upper bound we use the fact that for a hermitian operator $A$ we have $\vert A \vert = A_+ + A_-$ and
\begin{align}
    \Vert A\Vert_1 = \operatorname{tr}[\vert A\vert ] = \operatorname{tr}[A_+] + \operatorname{tr}[A_-].
\end{align}
Furthermore, we have of course $\operatorname{tr}[A] = \operatorname{tr}[A_+] -\operatorname{tr}[A_-]$. Together we get
\begin{align}
    \operatorname{tr}[A_+] = \frac{\Vert X\Vert_1 + \operatorname{tr}[A]}{2}.
\end{align}
Inserting $A = \sigma s - \rho$, yields with the triangle inequality $\Vert \sigma s - \rho \Vert_1 \leq \Vert \sigma s \Vert_1 + \Vert \rho \Vert_1 \leq s + 1$
\begin{align}
    \operatorname{tr}^+[\sigma s - \rho] = \frac{\Vert \sigma s -\rho\Vert_1 + s -1}{2} \leq \frac{s+1 + s -1}{2} = s.
\end{align}
Similarly with $\Vert A\Vert_1 \geq \vert \operatorname{tr}[A]\vert = \vert s-1\vert$
\begin{align}
    \operatorname{tr}^+[\sigma s - \rho] = \frac{\Vert \sigma s -\rho\Vert_1 + s -1}{2} \geq \frac{\vert s -1 \vert + (s-1)}{2} \geq s-1.
\end{align}
\end{proof}
\section{Proof of \texorpdfstring{\autoref{lem:convergence_trapezoid}}{}}\label{proof_convergence}
\begin{proof}
  By convexity, the one‐sided derivatives
  \begin{align}
    g^\prime_+(t_k)
    =\lim_{h\to0^+}\frac{g(t_k+h)-g(t_k)}h,
    \quad
    g^\prime_-(t_{k+1})
    =\lim_{h\to0^+}\frac{g(t_{k+1})-g(t_{k+1}-h)}h
  \end{align}
  exist and satisfy $g^\prime_+(t_k)\leq m_k\leq g^\prime_-(t_{k+1})$. Hence for $t\in[t_k,t_{k+1}]$
  \begin{equation}
  \begin{aligned}
    &S_k(t)-g(t)
    \leq  \min\Bigl\{
      \underbrace{S_k(t)-\bigl(g(t_k)+g^\prime_+(t_k)(t-t_k)\bigr)}_{v(t)},\\
      &\hspace{4cm}
      \underbrace{S_k(t)-\bigl(g(t_{k+1})-g^\prime_-(t_{k+1})(t_{k+1}-t)\bigr)}_{w(t)}
    \Bigr\},
  \end{aligned}
  \end{equation}
  where
  \begin{align}
    v(t)=(m_k - g^\prime_+(t_k))(t-t_k),
    \quad
    w(t)=(g^\prime_-(t_{k+1})-m_k)(t_{k+1}-t).
  \end{align}
  Solve $v(t)=w(t)$:
  \begin{align*}
    (m_k - g^\prime_+(t_k))t - (m_k - g^\prime_+(t_k))t_k
    &= (g^\prime_-(t_{k+1})-m_k)t_{k+1}
      - (g^\prime_-(t_{k+1})-m_k)t,\\
    (m_k - g^\prime_+(t_k) + g^\prime_-(t_{k+1})-m_k)t
    &= (g^\prime_-(t_{k+1})-m_k)t_{k+1}
      + (m_k - g^\prime_+(t_k))t_k,\\
    \bigl(g^\prime_-(t_{k+1})-g^\prime_+(t_k)\bigr)t^*
    &= (g^\prime_-(t_{k+1})-m_k)t_{k+1}
      + (m_k - g^\prime_+(t_k))t_k,\\
    t^*
    &= \frac{
         (g^\prime_-(t_{k+1})-m_k)t_{k+1}
         + (m_k-g^\prime_+(t_k))t_k
       }{
         g^\prime_-(t_{k+1})-g^\prime_+(t_k)
       }.
  \end{align*}
  It follows that
  \begin{equation}
  \begin{aligned}
    t^*-t_k
    = \frac{g^\prime_-(t_{k+1})-m_k}{g^\prime_-(t_{k+1})-g^\prime_+(t_k)}(t_{k+1}-t_k),\\
    t_{k+1}-t^*
    = \frac{m_k-g^\prime_+(t_k)}{g^\prime_-(t_{k+1})-g^\prime_+(t_k)}(t_{k+1}-t_k).
  \end{aligned}
  \end{equation}
  Integrating the two triangles,
  \begin{equation}
  \begin{aligned}
    \int_{t_k}^{t_{k+1}}\min\{v,w\}
    &=\int_{t_k}^{t^*}v(t)\,dt + \int_{t^*}^{t_{k+1}}w(t)\,dt \\
    &= \tfrac12(t^*-t_k)v(t^*) + \tfrac12(t_{k+1}-t^*)w(t^*).
  \end{aligned}
  \end{equation}
  At $t^\star$ one has
\begin{align}
  v(t^*)=w(t^*)
  = \frac{(m_k - g^\prime_+(t_k))(g^\prime_-(t_{k+1})-m_k)}{g^\prime_-(t_{k+1})-g^\prime_+(t_k)}(t_{k+1}-t_k).
\end{align}
\noindent Finally, by the elementary inequality
\begin{align}
  ab \le \frac{(a+b)^2}{4}\qquad\text{(a consequence of }(a-b)^2\ge 0\text{)},
\end{align}
applied with
\begin{align}
  a:=m_k - g^\prime_+(t_k),\qquad b:=g^\prime_-(t_{k+1})-m_k,\qquad
  a+b=g^\prime_-(t_{k+1})-g^\prime_+(t_k),
\end{align}
we obtain
\begin{align}
  (m_k - g^\prime_+(t_k))(g^\prime_-(t_{k+1})-m_k)
  \le
  \Bigl(\tfrac{g^\prime_-(t_{k+1})-g^\prime_+(t_k)}{2}\Bigr)^{\!2}.
\end{align}
Therefore, for $h_k=t_{k+1}-t_k$,
\begin{equation}
\begin{aligned}
\int_{t_k}^{t_{k+1}} \! S_k(t) - g(t)\, dt &\leq\tfrac12\Bigl(\tfrac{g^\prime_-(t_{k+1}) -m_k}{g^\prime_-(t_{k+1}) - g^\prime_+(t_k)}h_k\Bigr)
            \Bigl(\tfrac{(m_k - g^\prime_+(t_k))(g^\prime_-(t_{k+1}) -m_k)}{g^\prime_-(t_{k+1}) - g^\prime_+(t_k)}h_k\Bigr)\\
  &\quad+\tfrac12\Bigl(\tfrac{m_k - g^\prime_+(t_k)}{g^\prime_-(t_{k+1}) - g^\prime_+(t_k)}h_k\Bigr)
             \Bigl(\tfrac{(m_k - g^\prime_+(t_k))(g^\prime_-(t_{k+1}) -m_k)}{g^\prime_-(t_{k+1}) - g^\prime_+(t_k)}h_k\Bigr)\\
  &=\frac{(m_k - g^\prime_+(t_k))(g^\prime_-(t_{k+1}) -m_k)}{2\bigl(g^\prime_-(t_{k+1}) - g^\prime_+(t_k)\bigr)}h_k^2 \\
  &\le
  \frac{1}{2\bigl(g^\prime_-(t_{k+1}) - g^\prime_+(t_k)\bigr)}
  \Bigl(\frac{g^\prime_-(t_{k+1}) - g^\prime_+(t_k)}{2}\Bigr)^{\!2}h_k^2\\
  &=\frac{g^\prime_-(t_{k+1}) - g^\prime_+(t_k)}{8}\,h_k^2. 
\end{aligned}
\end{equation}
  Summing over all $k=1,\dots,r-1$ yields exactly the desired result:
\begin{align}
\sum_{k=1}^{r-1}\int_{t_k}^{t_{k+1}}S_k(t)\,dt -\int_\mu^\lambda g(t)\,dt
\leq
\sum_{k=1}^{r-1}\frac{(t_{k+1}-t_k)^2}{8}\bigl(g^\prime_-(t_{k+1})-g^\prime_+(t_k)\bigr).
\end{align}

\end{proof}
\section{Proof of \texorpdfstring{\autoref{lem:lower_bounds_at_least_upper_bounds}}{}}\label{appendix_proof_lemma_2}

\begin{proof}
Set $\xi(s)\coloneqq g_+'(s)$. For convex $g$, $\xi$ exists for all $s\in(a,b)$, is nondecreasing, and
$g_+'(a)\le \xi(s)\le g_-'(b)$.
By \cite[Thm.~1.5.2]{Niculescu2025}, for every $s\in[a,b]$,
\begin{equation}\label{eq:rep-no-mu}
g(s)=g(a)+\int_a^s \xi(u)\,du .
\end{equation}

Fix $s^\star\in[a,b]$ and a supporting slope $m\in\partial g(s^\star)=[g_-'(s^\star),g_+'(s^\star)]$.
Let $T_{a,b}(s)=g(s^\star)+m(s-s^\star)$ and $h(s)\coloneqq g(s)-T_{a,b}(s)\ge 0$.
Using \eqref{eq:rep-no-mu} at $s$ and $s^\star$,
\begin{align}
h(s)=\int_{s^\star}^s \bigl(\xi(u)-m\bigr)\,du .
\end{align}
Integrating $h$ over $[a,b]$ and splitting at $s^\star$, Fubini-Tonelli gives
\begin{align}
\int_a^b h(s)\,ds
&= \int_a^{s^\star}\int_s^{s^\star}\!\bigl(m-\xi(u)\bigr)\,du\,ds
   +\int_{s^\star}^b\int_{s^\star}^s\!\bigl(\xi(u)-m\bigr)\,du\,ds \notag\\
&= \int_a^{s^\star} \bigl(m-\xi(u)\bigr)(u-a)\,du
  +\int_{s^\star}^b \bigl(\xi(u)-m\bigr)(b-u)\,du. \label{eq:split-m-no-mu}
\end{align}

Decompose on $(a,s^\star]$ and $[s^\star,b)$:
\begin{align}
m-\xi(u)=\bigl(m-\xi(s^\star)\bigr)+\bigl(\xi(s^\star)-\xi(u)\bigr),\qquad
\xi(u)-m=\bigl(\xi(u)-\xi(s^\star)\bigr)+\bigl(\xi(s^\star)-m\bigr).
\end{align}
Plugging this into \eqref{eq:split-m-no-mu} yields
\begin{align*}
\int_a^b \!\bigl(g-T_{a,b}\bigr)\,ds
&= \int_a^{s^\star} \bigl(\xi(s^\star)-\xi(u)\bigr)(u-a)\,du
  +\int_{s^\star}^{b} \bigl(\xi(u)-\xi(s^\star)\bigr)(b-u)\,du \\
&\quad +\frac{(s^\star-a)^2-(b-s^\star)^2}{2}\,\bigl(m-\xi(s^\star)\bigr).
\end{align*}

To obtain an upper bound that is \emph{uniform in $m\in[g_-'(s^\star),g_+'(s^\star)]$}, treat the last term by cases:

\begin{itemize}
\item If $(s^\star-a)\ge(b-s^\star)$, then the coefficient of $m-\xi(s^\star)$ is nonnegative while $m-\xi(s^\star)\le 0$; hence this term is $\le 0$ and can be dropped.
\item If $(s^\star-a)\le(b-s^\star)$, then the coefficient is nonpositive, so the last term is maximized by $m=g_-'(s^\star)$, giving
\begin{align}
\frac{(s^\star-a)^2-(b-s^\star)^2}{2}\,\bigl(m-\xi(s^\star)\bigr)
\;\le\; \frac{(b-s^\star)^2-(s^\star-a)^2}{2}\,\bigl(\xi(s^\star)-g_-'(s^\star)\bigr)
\;\le\; \frac{(b-s^\star)^2}{2}\,\bigl(\xi(s^\star)-g_-'(s^\star)\bigr).
\end{align}
\end{itemize}

For the remaining two integrals, use only monotonicity of $\xi$:
\begin{align}
0\le \xi(s^\star)-\xi(u)\le \xi(s^\star)-\xi(a)\quad (u\in[a,s^\star]),\qquad
0\le \xi(u)-\xi(s^\star)\le \xi(b)-\xi(s^\star)\quad (u\in[s^\star,b]).
\end{align}
Hence
\begin{align}
\int_a^{s^\star} \bigl(\xi(s^\star)-\xi(u)\bigr)(u-a)\,du
\le \bigl(\xi(s^\star)-\xi(a)\bigr)\!\int_a^{s^\star}\!(u-a)\,du
= \frac{(s^\star-a)^2}{2}\,\bigl(\xi(s^\star)-\xi(a)\bigr),
\end{align}
and similarly
\begin{align}
\int_{s^\star}^{b} \bigl(\xi(u)-\xi(s^\star)\bigr)(b-u)\,du
\le \frac{(b-s^\star)^2}{2}\,\bigl(\xi(b)-\xi(s^\star)\bigr).
\end{align}
Using $\xi(a)=g_+'(a)$, $\xi(s^\star)=g_+'(s^\star)$, $\xi(b)=g_-'(b)$, we conclude
\begin{align}
\int_a^b \!\bigl(g-T_{a,b}\bigr)\,ds
\;\le\;
\frac12\,(s^\star-a)^2\bigl(g_+'(s^\star)-g_+'(a)\bigr)
+\frac12\,(b-s^\star)^2\bigl(g_-'(b)-g_-'(s^\star)\bigr).
\end{align}

Furthermore, we have
$g_+'(s^\star)-g_+'(a)\le g_-'(b)-g_+'(a)$ and
$g_-'(b)-g_-'(s^\star)\le g_-'(b)-g_+'(a)$, and then take
$\max\{(s^\star-a)^2,(b-s^\star)^2\}$.
Finally, by taking $s^\star=(a+b)/2$, for which
$\max\{(s^\star-a)^2,(b-s^\star)^2\}=(b-a)^2/4$ and Hölder's inequality for $w\ge 0$ and $L_{a,b}=\sup_{[a,b]} w$, we conclude the desired result.
\end{proof}

\section{SDP formulations for Standard Task}\label{apendix:sdp_formulations}
Material 
The method presented here is able to solve tasks which have one of the following properties
\begin{enumerate}
    \item It is a minimisation over a positive weighted sum of relative entropies 
    \begin{align*}
       \text{min} \sum_i \zeta_i &D(\rho^{(i)} \Vert \sigma^{(i)}) \\
       \operatorname{s.t.} \ &\text{semidefinite constraints on the states}
    \end{align*}
    i.e. $\zeta_i \geq 0$ for all $i$
    \item each state within each relative entropy is linear in the sdp variables, i.e. something like $\rho^{(i)} = \Phi(\tilde{\rho}^{(i)})$ whereby $\tilde{\rho}^{(i)}$ is the actual SDP variable. Exemplary $\Phi$ could be the channel that tensorises just a state or a measurement.  
\end{enumerate}
Therefore we can estimate sums of e.g. the following summands 

\begin{enumerate}
    \item 
For the entropy of a state $\rho$ under linear constraints we can maximise $H(\rho) = \log d - D(\rho \Vert \mathds{1}/d)$ under linear constraints if we equivalently minimise
\begin{equation*}
\begin{aligned}
    \inf_{\rho} \ &D(\rho \Vert \mathds{1}/d) \\
     \operatorname{s.t.} & \  \; h_i(\rho)\geq 0 \quad i=1,\dots,n \\ 
     &\mu \mathds{1} \leq \rho \leq \lambda \mathds{1}  \\
     &\rho \in \mathcal{S}(\mathcal{H})
\end{aligned}
\qquad
\Rightarrow
\qquad 
\begin{aligned}
    \inf_{\rho,\nu_k} &\sum_k \tr[\nu_k] + \log \lambda + 1 - \lambda\\ 
   \operatorname{s.t.}
    & \  \; h_i(\rho)\geq 0 \quad i=1,\dots,n \\ 
    &   \nu_k \geq \gamma_k \rho + \delta_k \mathds{1}  \quad \forall k \\ 
    &\mu \mathds{1} \leq \rho \leq \lambda \mathds{1}  \\
     &\rho \in \mathcal{S}(\mathcal{H}) \quad \nu_k\geq 0
\end{aligned}
\end{equation*}
\item For the conditional entropy of a state $\rho$ under linear constraints we can maximise 
\begin{align*}
    H(A\vert B)_{\rho_{AB}} = \log d_A - D(\rho_{AB} \Vert \mathds{1}_A/d_A \otimes \rho_B)    
\end{align*}
under linear constraints if we equivalently minimise
\begin{equation*}
\begin{aligned}
    \inf_{\rho} \ &D(\rho_{AB} \Vert \mathds{1}/d_A \otimes \rho_B) \\
     \operatorname{s.t.} & \  \; h_i(\rho_{AB})\geq 0 \quad i=1,\dots,n \\ 
     &\mu \mathds{1}/d_A\otimes \rho_B \leq \rho_{AB} \leq \lambda \mathds{1}/d_A\otimes \rho_B  \\
     &\rho_{AB} \in \mathcal{S}(\mathcal{H}_A \otimes \mathcal{H}_B)
\end{aligned}
\qquad
\Rightarrow
\qquad 
\begin{aligned}
    \inf_{\rho,\nu_k} &\sum_k \tr[\nu_k] + \log \lambda + 1 - \lambda\\ 
    \operatorname{s.t.} & \  \; h_i(\rho_{AB}) \geq 0 \quad i=1,\dots,n \\ 
     &\mu \mathds{1}/d_A\otimes \rho_B \leq \rho_{AB} \leq \lambda \mathds{1}/d_A\otimes \rho_B  \\
     &\rho_{AB} \in \mathcal{S}(\mathcal{H}_A \otimes \mathcal{H}_B) \quad \nu_k\geq 0.
\end{aligned}
\end{equation*}
\end{enumerate}

\section{Exemplary Application in Quantum Shannon Theory}\label{sec:quantum_shannon}

We consider exemplary the task of calculating the entanglement-assisted classical capacity similar to \cite[3.2]{Fawzi_2018} of the amplitude damping channel. This section is intended to demonstrate the strength and diversity of our approach rather than to maximising out its numerical limitations.

We consider the amplitude damping channel in its basic form with Kraus-representation 
\begin{align*}
   K_1 \coloneqq  \begin{pmatrix}
        1 & 0 \\
        0 & \sqrt{1-p}
    \end{pmatrix} \quad K_2 \coloneqq \begin{pmatrix}
        0 & \sqrt{p}\\
        0 & 0.
    \end{pmatrix}
\end{align*}
As shown in the \href{https://github.com/gereonkn/relative-entropy-optimization.git}{repository} it is easy to find an isometry that implements the dilation $U$ of the amplitude channel. Reference \cite{Bennett2002} has shown that the entanglement-assisted classical capacity can be formulated as the following optimisation task
\begin{align*}
    C(\mathcal{A}_p) \coloneqq \max_{\sigma \in \mathcal{S}(\mathcal{H})} I(\sigma,\mathcal{A}_p),
\end{align*}
whereby $\mathcal{A}_p$ is the amplitude damping channel with damping parameter $p$ and $I$ the mutual information. For our specific case, this yields 
\begin{align*}
    I(\sigma,\mathcal{A}_p) = H(B\vert E)_{U\sigma U^\dagger} + H(B)_{U\sigma U^\dagger}.
\end{align*}
Particularly we can rewrite this expression for $\rho_{BE} \coloneqq U \sigma U^\dagger$ to be 
\begin{align*}
    I(\rho_{BE},\mathcal{A}_p) = \log (d_B^2) - D(\rho_{BE} \Vert \mathds{1}_B/d_B \otimes \rho_E) - D(\rho_B \Vert \mathds{1}_B/d_B).
\end{align*}
That means optimising $I$ is equivalent to minimising 
\begin{align*}
    \min_{\rho = U\sigma U^\dagger} D(\rho_{BE} \Vert \mathds{1}_B/d_B \otimes \rho_E) + D(\rho_B \Vert \mathds{1}_B/d_B).
\end{align*}
This problem can be easily implemented if we just add our approach two times from \eqref{eq:relaxation_lower_bound} and \eqref{eq:relaxation_upper_bound} with two sets of variables $\tau$ but just one $\rho$. 
\begin{figure}
    \centering
    \begin{tikzpicture}
\begin{axis}[width=10cm,height=6cm,
    xlabel = damping parameter $p$,
    ylabel = capacity]
\addplot[color=PineGreen]%
table{results/values_amplitude_fin.txt};
\addlegendentry{entanglement-assisted classical capacity}
\end{axis}
\end{tikzpicture}
    \caption{The plot shows exemplarily the entanglement assisted capacity of the amplitude damping channel on a qubit in dependence of the damping parameter. The plot is similar to figure 1 in \cite{Fawzi_2018}.}
    \label{fig:entanglemen_assisted_capacity}
\end{figure}

\section{Entanglement Measures}\label{sec:entanglement_measures}

This section is devoted to a "proof-of-principle" of our technique for estimating entanglement measures. We aim to estimate for a fixed $\rho_{AB} \in \mathcal{S}(\mathcal{H}_A \otimes \mathcal{H}_B)$ the following optimisation problem
\begin{align}\label{eq:relative_entropy_of_entanglement}
    \min_{\sigma_{AB} \in \text{SEP}(A:B)} D(\rho_{AB} \Vert \sigma_{AB}).
\end{align}
We denote to be $\text{SEP}(A:B)$ to be the set of all separable states over the bipartition $\mathcal{H}_A \otimes \mathcal{H}_B$. The quantity in \eqref{eq:relative_entropy_of_entanglement} is often called \emph{relative entropy of entanglement} \cite{Piani_2009} and has a significant meaning in entanglement theory.

Nevertheless, there are SDP-hierarchies to solve \eqref{eq:relative_entropy_of_entanglement} if the functional would be a usual linear SDP \cite{Doherty_2005}. Combining our linearisation and a hierarchy for the set of separable states could therefore yield reasonable lower bounds (the hierarchies are optimising over e.g. $n$-extendable states and therefore relax the problem in terms of lower bounds and we are providing in particular lower bounds as well). 

We would like to repeat that the aim of this appendix is not to exhaust the numerical possibilities, but rather to demonstrate the versatility of our approach by way of example. Thus, we use the fact that for two qubits the PPT criterium is sufficient \cite{Horodecki_1996} and show the relative entropy of entanglement for an entangled state on two qubits mixed with white noise. 

\begin{figure}
    \centering
    \begin{tikzpicture}
\begin{axis}[width=10cm,height=6cm,
    xlabel = alpha,
    ylabel = relative entropy of entanglement]
\addplot[color=PineGreen]%
table{results/relative_entropy_entanglement.txt};
\addlegendentry{relative entropy of entanglement}
\end{axis}
\end{tikzpicture}
    \caption{The plot shows the relative entropy of entanglement on a pair of qubits here with PPT criterium in dependence with a maximally entangled state mixed with white noise. The amount of white noise is calculated with a parameter alpha.}
    \label{fig:relative_entropy_entanglement}
\end{figure}

\end{widetext}
\end{document}